\documentclass[12pt]{article}
\pdfoutput=1
\usepackage{graphicx}
\usepackage{jheppub}
\usepackage{bbold}
\usepackage[T1]{fontenc} % if needed
\usepackage{slashed}
\usepackage{color}
\usepackage{hyperref}
\renewcommand{\text}[1]{#1}

\newcommand{\be}{\begin{equation}}
\newcommand{\ee}{\end{equation}}
\newcommand{\ben}{\begin{displaymath}}
\newcommand{\een}{\end{displaymath}}
\newcommand{\bea}{\begin{eqnarray}}
\newcommand{\eea}{\end{eqnarray}}
\newcommand{\bean}{\begin{eqnarray*}}
\newcommand{\eean}{\end{eqnarray*}}
\newcommand{\nn}{\nonumber \\}
\newcommand{\ba}{\begin{array}}
\newcommand{\ea}{\end{array}}
\newcommand{\bi}{\begin{itemize}}
\newcommand{\ei}{\end{itemize}}

\def\a{\alpha}

\def\g{\gamma}
\def\G{\Gamma}

\def\G{\Gamma}
\def\g{\gamma}
\def\e{\epsilon}
\def\s{\sigma}
\def\e{\epsilon}

\def\m{\mu}

\DeclareMathOperator{\vol}{vol}

\title{\boldmath An alternative IIB embedding of F(4) gauged supergravity}

\author[a]{Jaehoon Jeong,}
\author[b]{\"Ozg\"ur Kelekci,}
\author[c]{Eoin \'O Colg\'ain\,}

\affiliation[a]{Center for Quantum Spacetime, Sogang University,
Seoul 121-741, Korea}
\affiliation[b]{Department of Physics and Graphene Research Institute, Sejong University, \\ Seoul 143-747, Korea}
\affiliation[c]{Departamento de F\'isica, Universidad de Oviedo, Oviedo 33007, Espa\~na}

\emailAdd{jhjeong@sogang.ac.kr}
\emailAdd{okelekci@sejong.edu}
\emailAdd{ocolgain@gmail.com}

\abstract{Through the construction of a complete non-linear Kaluza-Klein reduction ansatz from type IIB supergravity to Romans' F(4) gauged supergravity, we identify a recently discovered supersymmetric $AdS_6$ solution as the IIB uplift of the supersymmetric vacuum of Romans' theory. We present new IIB uplifts of a number of known solutions of Romans' theory and comment on supersymmetry in higher-dimensions where it is expected. }

\begin{document}
\begin{flushright}
{CQUeST--2013-0583\\FPAUO-13/02}
\end{flushright}
\maketitle

\flushbottom

\section{Introduction}
Late last year, we witnessed the identification of the first examples of supersymmetry preserving non-Abelian T-duality transformations \cite{Lozano:2012au, Itsios:2012zv, Itsios:2013wd} which, in one case \cite{Lozano:2012au}, led to the unexpected discovery of what may be regarded as a supersymmetric $AdS_6$ doppelg\"anger geometry in type II supergravity. To put this result into proper context, it is well over a decade since the only solution in this class was identified \cite{Brandhuber:1999np} in massive IIA supergravity \cite{Romans:1985tz} and recent reports were veering slowly towards uniqueness statements \cite{Passias:2012vp}\footnote{The absence of other supersymmetric vacua in the matter coupled theory \cite{D'Auria:2000ad} is touched upon in \cite{Karndumri:2012vh}.}. Against this backdrop, the purpose of this note is to unmask our doppelg\"anger as simply the supersymmetric vacuum of Romans' F(4) gauged supergravity \cite{Romans:1985tw}, but in a less familiar ten-dimensional guise.

To put Romans' theory in a historical context, recall that Nahm's 1978 classification of simple superalgebras  \cite{Nahm:1977tg} acted as the catalyst for the quest to identify supergravity theories with vacua invariant under the global symmetries of these algebras. Building on successes in the identification of supergravities with vacua invariant under OSp(8|4,R) \cite{de Wit:1982ig}, SU(2,2|4) \cite{Gunaydin:1984qu} and OSp(8$^*$|4) \cite{Pernici:1984xx}, one thread of this fascinating detective story ended in 1985 when the supergravity corresponding to the exceptional superalgebra F(4) was discovered. Romans' important observation was that a mass parameter for the two-index tensor of the $N=4$ theory \cite{Giani:1984dw} could be introduced leading to a gauged supergravity \cite{Romans:1985tw} with two $AdS_6$ vacua, one of which is supersymmetric. In a parallel development it was understood that all these supergravities were simply ten and eleven-dimensional supergravity reduced consistently on spheres \cite{s51,lpt,s53,s52,dewit,s41,Cvetic:1999un}.

In fact, as hinted at above, supersymmetry plays some r\^{o}le in consistent Kaluza-Klein (KK) dimensional reductions. In general, there is often no fundamental guiding principle in the construction of  KK reduction ans\"atze and the only recourse can be trial and error. However, sometimes a symmetry principle is at work, such as an existing symmetry of the equations of motion, e.g. T-duality \cite{Bergshoeff:1995as,OColgain:2011ng,Itsios:2012dc}, the presence of a $G$-structure \cite{Gauntlett:2009zw, Cassani:2010uw, Gauntlett:2010vu,Skenderis:2010vz, Liu:2010sa, OColgain:2010rg, Cassani:2011fu}, or when the internal space is a coset manifold \cite{Cassani:2009ck, Cassani:2010na,Bena:2010pr,Cassani:2012pj}. These situations aside, the identification of KK reductions remains a daunting exercise, but supersymmetry can offer valuable insights. Generalising conclusions drawn in \cite{Buchel:2006gb,Gauntlett:2006ai} and through the elucidation of further examples, it was conjectured in \cite{Gauntlett:2007ma} that gauging R-symmetries always leads to consistent KK reductions to lower-dimensional supergravities admitting AdS vacua. To test this conjecture further, \cite{Gauntlett:2007sm} exhibited an elegant example of this conjecture by showing that the Lin, Lunin, Maldacena (LLM) class \cite{LLM} of geometries\footnote{See \cite{OColgain:2010ev} for comments on the generality of the LLM geometries.} dual to SCFTs with R-symmetry $SU(2) \times U(1)$, can be reduced to Romans' five-dimensional $SU(2) \times U(1)$ gauged supergravity \cite{Romans:1985ps}.

Through the benefit of hindsight, we can now view the consistent KK reduction of massive IIA supergravity on $S^4$ \cite{Cvetic:1999un} to Romans' F(4) gauged supergravity \cite{Romans:1985tw} through the prism of this conjecture. Since the $AdS_6 \times S^4$ is warped \cite{Brandhuber:1999np},  the natural $SO(5)$ isometry is broken to $SO(4) \sim SU(2) \times SU(2)$, where only a single $SU(2)$ factor corresponds to the R-symmetry. This particular $SU(2)$ factor is then singled out through the writing of $S^3$ in terms of left-invariant one-forms \cite{Cvetic:1999un}. Then according to our conjecture \cite{Gauntlett:2007ma}, we should expect that gaugings of the R-symmetry lead to a theory with an $SU(2)$ gauge group and presumably the mass parameter comes along for the ride, resulting in a lower-dimensional massive gauged supergravity. Scouring the literature, one finds a single theory fitting this billing, namely Romans' F(4) gauged supergravity \cite{Romans:1985tw}. The point of this work is that now we have a new supersymmetric $AdS_6$ vacuum in type IIB \cite{Lozano:2012au} with the required $SU(2)$ R-symmetry manifest in an $S^2$ factor, so we can gauge the $S^2$ leading to the same result.

Together, the original reduction of Cveti\v{c} et. al \cite{Cvetic:1999un}, and the new embedding of Romans' theory in type IIB we present here, open up Romans' theory to the string theory community since  it is technically easier to find solutions via ansatz in lower-dimensions and then uplift. Indeed, in the past, we have seen supersymmetric domain walls \cite{Lu:1995hm}, solutions dual to twisted field theories \cite{Nunez:2001pt}, RG flows \cite{Gursoy:2002tx}, various black holes \cite{Chong:2004ce,Chow:2008ip} and more recently Lifshitz geometries \cite{Gregory:2010gx,Braviner:2011kz,Barclay:2012he} constructed directly in Romans' theory, before the connection to ten-dimensions was exploited. Here we emphasise that there is not just one uplift, but two\footnote{In fact, there are three and counting as the Abelian T-dual of \cite{Cvetic:1999un} will give another.}, so the number of uplifted solutions doubles.

\begin{figure}[h]
\label{fig:red}
\centering
\includegraphics[width=120mm]{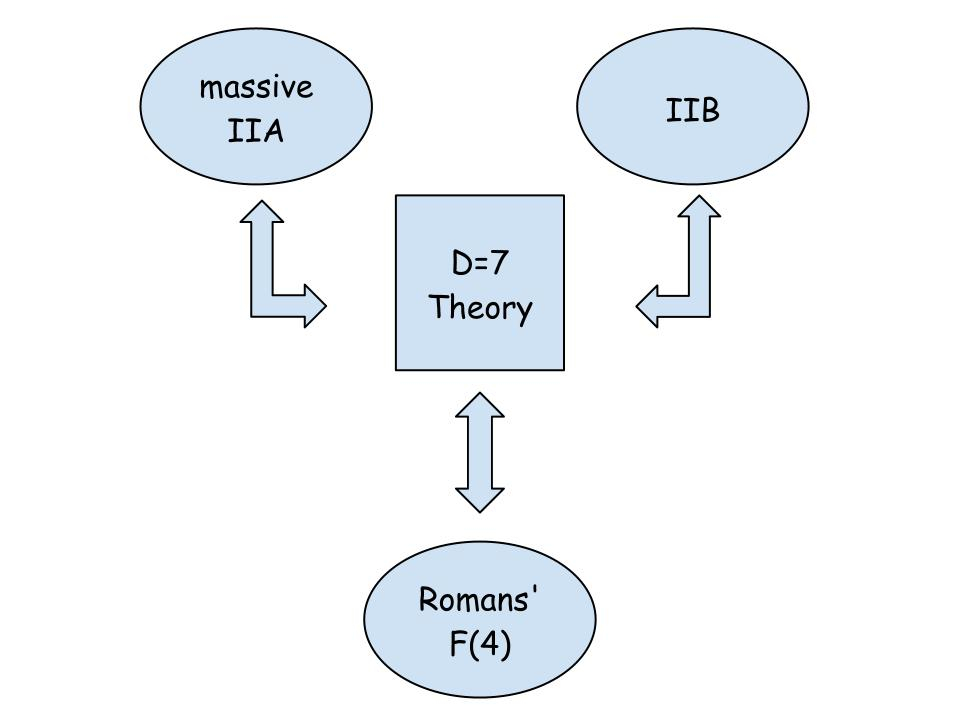}
\caption{The massive IIA reduction on $S^4$ can be decomposed into a reduction on $S^3$ (left arrow) to $D=7$ followed by a further reduction on the remaining angular coordinate of the $S^4$. In this paper we construct the alternative reduction from type IIB (right arrow) to $D=7$ leading to an embedding of Romans' theory in type IIB.}
\end{figure}

Last year also marked a small resurgence of interest in the AdS/CFT within the scope of five-dimensional theories. The strongly-coupled supersymmetric fixed-point theories pioneered in \cite{Seiberg:1996bd, Morrison:1996xf, Intriligator:1997pq} were revisited and quiver gauge theories dual to $AdS_6 \times S^4/\mathbb{Z}_n$ were constructed \cite{Bergman:2012kr}. Subsequently, the Higgs branch of the theories was probed by dual giants \cite{Bergman:2012qh}. Localization techniques also featured prominently: addressing global symmetry enhancement  \cite{Kim:2012gu}, an exact computation of the $S^5$ partition function of SCFTs dual to $AdS_6 \times S^4$ led to perfect agreement \cite{Jafferis:2012iv}, and finally a study of half-bps Wilson loops \cite{Assel:2012nf} was shown to match up with supersymmetric D4-brane probes at large $N$. In this setting, the question of whether this new $AdS_6$ solution has a bona fide CFT dual will be broached in \cite{todo}.\footnote{In particular,  we plan to make sense of the non-Abelian T-dual coordinate $r$ which will need to be compactified if one is to quantise fluxes and assign D-brane charges correctly. On the other hand, for small $r$, the T-dual geometry smoothly approaches $\mathbb{R}^3$. This important point is a key prerequisite for further discussion on the global properties of the uplifted IIB solutions which we have to yet show are globally well-behaved. We observe here that both the Abelian and non-Abelian T-dual of $AdS_6 \times S^4/\mathbb{Z}_2$ have curvature singularities at both end-points of the polar angle for $S^4/\mathbb{Z}_2$ and are thus more singular than the original geometry. }

However, back to the matter at hand. Key to our construction of a KK reduction ansatz will be non-Abelian T-duality, a transformation which was initially studied in  \cite{delaOssa:1992vc,Giveon:1993ai,Sfetsos:1994vz,Alvarez:1994np} and has gone through a particular  purple patch of late \cite{Sfetsos:2010uq, Lozano:2011kb, Itsios:2012dc, Itsios:2013wd, Lozano:2012au, Itsios:2012zv} leading to a greater understanding of solution generation in type II supergravity. To exploit this angle, we will construct a consistent KK reduction ansatz from type IIB supergravity to Romans' theory in two steps. We start by remarking that the original KK reduction from massive IIA \cite{Cvetic:1999un} can be broken up into an initial reduction on $S^3$ to seven-dimensions, followed by a subsequent reduction to six-dimensions. As non-Abelian T-duality simply transforms the $S^3$, we can view our construction as replacing the initial step of the massive IIA reduction on $S^3$ by an alternative reduction on the non-Abelian T-dual geometry, this time from type IIB supergravity. Thus, once we show in seven-dimensions that the equations of motion are the same, we can further reduce to six-dimensions to make the connection to Romans' theory. This philosophy is encapsulated in Figure \ref{fig:red}.

The structure of the rest of the paper runs thus. After reviewing Romans' theory in section \ref{sec:review}, in section \ref{sec:IIAred} we rewrite the reduction ansatz of \cite{Cvetic:1999un} in terms of seven-dimensional equations of motion, which will serve as ``target" equations.  In section \ref{sec:NS} we will deduce the NS sector of the non-Abelian T-dual and remark that one can use non-Abelian T-duality to derive this on the nose. We will at that point confirm that the dilaton equation from type IIB reduced to seven-dimensions agrees with our target equations, providing confirmation that we are on the right track to establish a connection at the level of the equations of motion in seven-dimensions. In section \ref{sec:IIBredRR}, we will complete the KK reduction ansatz by deducing the RR fluxes from a knowledge of the NS sector generated in section \ref{sec:NS}. Finally, plugging the ansatz into the type IIB equations of motion, we check that we recover the same equations of motion as in section \ref{sec:IIAred}, telling us that at both the seven-dimensional  and six-dimensional level, i.e. Romans' theory, the theories are the same. In section \ref{sec:solns} we focus our attention on uplifting various solutions to both massive IIA and type IIB, and where they are supersymmetric, we comment on the supersymmetry, before presenting our conclusions.

\section{Review of Romans' theory}
\label{sec:review}
We begin with a review of Romans' $D=6$ F(4) gauged supergravity \cite{Romans:1985tw}. More precisely, the theory of interest to us will be Romans $N= 4^+$ theory where both the  gauge coupling $g$ and the mass parameter $m$ are positive. This theory is then related to four other distinct theories for different values of the gauge coupling and mass parameter. Note that these are all described by the same Lagrangian and field content.

The theory consists of a graviton $e^{\a}_{\mu}$, three $SU(2)$ gauge potentials $A^i_{\mu}$, an Abelian potential $\mathcal{A}_{\mu}$, a two-index tensor gauge field $B_{\mu \nu}$, a scalar $\phi$, four gravitini $\psi_{\mu i}$ and four spin-$\tfrac{1}{2}$ fields $\chi_{i}$. The bosonic Lagrangian is
\bea
e^{-1}\,{\cal L}_6 &=& -\tfrac{1}{4} R +\tfrac{1}{2} (\partial \phi)^2 - \tfrac{1}{4} e^{-\sqrt{2} \phi} \left( \mathcal{H}^2 + (F^i)^2 \right) + \tfrac{1}{12} e^{2 \sqrt{2} \phi} G^2 + V  \nn
&-& \tfrac{1}{8}  \e^{\mu \nu \rho \sigma \tau \kappa} B_{\mu \nu} \left( \mathcal{F}_{\rho \sigma} \mathcal{F}_{\tau \kappa} + m B_{\rho \sigma} \mathcal{F}_{\tau \kappa} + \tfrac{1}{3} m^2 B_{\rho \sigma} B_{\tau \kappa} + F^i_{\rho \sigma} F^i_{\tau \kappa} \right),
\label{romansmassive}
\eea
where the potential $V$ is
\be
V = \tfrac{1}{8} \left( g^2 e^{\sqrt{2} \phi} + 4 g m e^{-\sqrt{2} \phi} - m^2 e^{-3 \sqrt{2} \phi}\right),
\ee
and, in addition, $e$ is the determinant of the vielbein, $g$ is the $SU(2)$ coupling constant and $m$ is the mass associated with $B_{\mu \nu}$. The field strengths in the action (\ref{romansmassive}) may be expressed as\footnote{Throughout we use the notation $\omega^2 \equiv \omega_{i_1\dots i_p} \omega^{i_1 \dots i_p}$ and  $(\omega^2)_{\mu \nu} = \omega_{\mu \s_1\dots \s_{p-1}} \omega_{\nu}^{~ \s_1\dots \s_{p-1}}$ for $p$-forms.}
\bea
\mathcal{F}_{\mu \nu} &\equiv& \partial_{\mu} \mathcal{A}_{\nu} - \partial_{\nu} \mathcal{A}_{\mu}, \nn
F^i_{\mu \nu} &\equiv& \partial_{\mu} A^i_{\nu} - \partial_{\nu} A^i_{\mu} + g \e_{ijk} A^j_{\mu} A^k_{\nu}, \nn
G_{\mu \nu \rho} &\equiv& 3 \partial_{[\mu} B_{\nu \rho]}, \nn
\mathcal{H}_{\mu \nu} &\equiv& \mathcal{F}_{\mu \nu} + m B_{\mu \nu}.
\eea
We observe that the Lagrangian enjoys a global symmetry of the form
 \be
\phi \rightarrow \phi + \sqrt{2} \log \alpha, \quad \mathcal{A}_{\mu} \rightarrow \alpha \mathcal{A}_{\mu}, \quad A^i_{\mu} \rightarrow \a A^i_{\mu}, \quad B_{\mu \nu} \rightarrow \a^{-2} B_{\mu \nu}
 \ee
 provided the parameters are also rescaled
 \be
 g \rightarrow \a^{-1} g, \quad m \rightarrow \a^3 m.
 \ee
 This global symmetry may be exploited to set the scalar to zero whenever it is a constant.

 As the theme of this paper is dimensional reductions from type II supergravity, it is useful to re-express Romans' theory in a form that permits an immediate uplift on $S^4$ to massive IIA supergravity \cite{Romans:1985tz}.  The lower-dimensional theory  in the language of differential forms of \cite{Cvetic:1999un} may be expressed as
\bea
\label{cveticaction}
\tilde{\mathcal{L}}_6 &=& \tilde{R} * \mathbb{1} - \tfrac{1}{2} * d \tilde{\phi} \wedge d \tilde{\phi} - \tilde{g}^2 \left( \tfrac{2}{9} e^{\frac{3}{\sqrt{2}} \tilde{\phi}} - \tfrac{8}{3} e^{\frac{1}{\sqrt{2}} \tilde{\phi}} - 2 e^{\frac{1}{-\sqrt{2}} \tilde{\phi}} \right) * \mathbb{1} \nn
&-& \tfrac{1}{2} e^{-\sqrt{2} \tilde{\phi} } * F_{(3)} \wedge F_{(3)} - \tfrac{1}{2} e^{\frac{1}{\sqrt{2}} \tilde{\phi}} \left( * F_{(2)} \wedge F_{(2)} + * \tilde{F}_{(2)}^i \wedge \tilde{F}_{(2)}^i\right) \\
&-& A_{(2)} \wedge \left(\tfrac{1}{2} d A_{(1)} \wedge d A_{(1)} + \tfrac{1}{3} \tilde{g} A_{(2)} \wedge d A_{(1)} + \tfrac{2}{27} \tilde{g}^2 A_{(2)} \wedge A_{(2)}  + \tfrac{1}{2} \tilde{F}^i_{(2)} \wedge \tilde{F}^{i}_{(2)}\right),  \nonumber
\eea
where we have defined the field strengths
\bea
F_{(3)} &=& d A_{(2)}, \nn
F_{(2)} &=& d A_{(1)} + \tfrac{2}{3} \tilde{g} A_{(2)}, \nn
\tilde{F}^i_{(2)} &=& d \tilde{A}^i_{(1)} + \tfrac{1}{2} \tilde{g} \e_{ijk} \tilde{A}^j_{(1)} \wedge \tilde{A}^k_{(1)}.
\eea
Tildes have been added where necessary to differentiate fields from the earlier notation of Romans (\ref{romansmassive}). These two actions can then be reconciled through the following redefinitons
\bea
&&\tilde{g}_{\mu \nu} = - g_{\mu \nu}, \quad \tilde{\phi} - 2 \tilde{\phi}_0 = - 2 \phi, \nn
&&e^{2 \sqrt{2} \tilde{\phi}_0} = 3 m g^{-1}, \quad \tilde{g} = \tfrac{1}{2} (3 m g^3)^{1/4}, \nn
&&\tfrac{1}{2} e^{1/ \sqrt{2} \tilde{\phi}_0} \tilde{F}^i_{(2)} = F^{i}, \quad  \tfrac{1}{2} e^{-\sqrt{2} \tilde{\phi}_0} F_{(3)} = G_3.
\label{cveticconn}
\eea
Observe here that the signature of the metric changes. The scalar also gets rescaled and shifted by a constant while the single gauge coupling parameter $\tilde{g}$ of \cite{Cvetic:1999un} may be recast in terms of the two parameters of Romans' theory. For brevity here we omit details of the KK reduction ansatz \cite{Cvetic:1999un} as the focus of the next section will be rewriting it in a $D=7$ guise.

\section{Reduction from IIA}
\label{sec:IIAred}
As mentioned earlier, the main thrust of this work is to show that Romans' F(4) gauged supergravity can be embedded in type IIB supergravity so that the supersymmetric vacuum in six-dimensions corresponds to the recently discovered supersymmetric $AdS_6$ solution of type IIB supergravity presented in \cite{Lozano:2012au}\footnote{The supersymmetric $AdS_5$ non-Abelian T-dual presented in \cite{Itsios:2013wd} reduces using the ansatz of \cite{Gauntlett:2006ai} to minimal $D=5$ gauged supergravity.}.
While we could work explicitly with the KK reduction ansatz of \cite{Cvetic:1999un}, as expressions are involved and our interest is effectively a non-Abelian T-duality transformation affecting only an internal $S^3$, in this section we rewrite the reduction of \cite{Cvetic:1999un} in terms of the equations of motion defining a particular $D=7$ theory. This theory can be further reduced to $D=6$ to recover the work of Romans.

Working in $D=7$ also facilitates contact with the reduction ansatz of  \cite{Itsios:2012dc}. In \cite{Itsios:2012dc} the ansatz considered involved a round $S^3$ without $SU(2)$ gauging. So, the space-time is assumed to be of the form
\be
ds^2 = ds^2(M_7) + e^{2A} ds^2(S^3),
\ee
where the warp factor $A$ is a scalar living on $M_7$ and we also have the following RR fluxes
\bea
&& F_0 = m \ ,
\nonumber\\
&& F_2 = G_2\ ,
\\
&& F_4 = G_4 + G_1\wedge {\rm Vol}(S^3)\ ,
\label{formsIIA}
\nonumber
\eea
and an additional $B$-field with field strength that has only components on the space-time $M_7$. The dilaton $\Phi$ is, like $A$, simply a scalar which depends on the coordinates of $M_7$.

Given a solution to massive IIA of the above form, we know that one can generate a non-Abelian T-dual and since simultaneous consistent reductions to the same $D=7$ theory exist from the both the original and T-dual geometries \cite{Itsios:2012dc}, we can deduce that the equations of motion get mapped.  The further observation then is that the reduction ansatz of \cite{Cvetic:1999un} fits into this template once we truncate out the $SU(2)$ gauge-fields. Therefore, any solution to Romans' F(4) supergravity \textit{without} $SU(2)$ gauge fields can be uplifted to type IIB supergravity on the non-Abelian T-dual. To stress this point further, this means that the supersymmetric vacuum aside \cite{Lozano:2012au}, a host of solutions, such as time-dependent D-branes \cite{Minamitsuji:2012if}, AdS solitons \cite{Horowitz:1998ha, Haehl:2012tw}, holographic RG flows \cite{Gursoy:2002tx, Karndumri:2012vh}, Kerr-AdS black holes \cite{Gibbons:2004js, Chen:2006xh} and the non-supersymmetric vacuum of Romans' theory \cite{Romans:1985tw} can be regarded as both solutions to massive IIA and type IIB supergravity.

Now to reinstate the $SU(2)$ gauge-fields and accommodate the full reduction ansatz of \cite{Cvetic:1999un}, we simply have to make the following changes to the reduction ansatz %\textcolor{red}{Here I am rescaling $e^{A} \rightarrow 2 e^{A}, G_1 \rightarrow 8 G_1$}:
\bea
\label{KKansatz}
ds^2 &=& ds^2(M_7) + e^{2 A} \sum_{i=1}^3 (\s^i - A^i)^2,  \nn
F_0 &=& m, \nn
F_2 &=& G_2 \nn
F_4 &=& G_4 + G_1 \wedge h_1 \wedge h_2 \wedge h_3 + h_i \wedge H^{i}_{3} + \tfrac{1}{2} \e_{ijk} H^i_{2} \wedge h_j \wedge h_k,
\eea
where $A^i, H_2^i$ and $H_3^i$ are additional one, two and three-forms with legs on $M_7$ and carrying $SU(2)$ indices, $\s^i$ are left-invariant one-forms on $S^3$ satisfying $d \s^i = - \frac{1}{2} \e_{ijk} \s^j \wedge \s^k$ and $h_i = \s^i - A^i$.
An explicit expression for these one-forms is
\bea
\s^1 &=&  \sin \phi d \theta - \cos \phi \sin \theta d \psi, \quad
\s^2 = \cos \phi d \theta + \sin \phi \sin \theta d \psi, \quad
\s^3 = d \phi + \cos \theta d \psi. \nonumber
\eea

In terms of the left-invariant one-forms, the metric on $S^3$,  normalised so that $R_{ij} = 2 g_{ij}$, takes the form:
\be
\label{s3}
ds^2(S^3) = \frac{1}{4} \left[ (\s^1)^2 + (\s^2)^2 + (\s^3)^2 \right],
\ee
so comparison with our ansatz reveals that the internal space is normalised so that $R_{ij} = \tfrac{1}{2} g_{ij}$. The choice of normalisations follows \cite{Sfetsos:2010uq, Itsios:2012dc} and simplifies consistency checks. Immediately, one can confirm that the original KK reduction ansatz \cite{Itsios:2012dc} is recovered when $A^i = H^{i}_{3} = H^{i}_{2} = 0$.

While we have not deformed the two-form field strength $F_2$ and it is obvious that one could consider greater generality, our choice of ansatz is motivated so that it the bare minimum covering the KK reduction ansatz of \cite{Cvetic:1999un}, modulo one distinction that we are working in string frame, so a rescaling of the metric is required.

To aid future consistency checks, we now relate the above fields to those appearing in \cite{Cvetic:1999un}. After rescaling the metric accordingly, direct comparison requires the following rewriting of our fields in terms of the notation of Cveti\v{c} et al.
\bea
\label{connect}
ds^2(M_7) &=&  X^{-\frac{1}{2}} s^{-\frac{1}{3}} \Delta^{\frac{1}{8}} \left[ \Delta^{\frac{3}{8}} ds^2_6 + 2 \tilde{g}^{-2} \Delta^{\frac{3}{8}} X^2 d \xi^2 \right] , \nn
e^{A} &=&  \tfrac{1}{\sqrt{2}} \tilde{g}^{-1} X^{-3/4} s^{-1/6} \Delta^{-1/4} c,\nn
A^i &=& \tilde{g} \tilde{A}^i_{(1)}, \nn
e^{\Phi} &=& s^{-5/6} \Delta^{1/4} X^{-5/4}, \nn
H &=& s^{2/3} F_{(3)} + \tilde{g}^{-1} s^{-1/3} c F_{(2)} \wedge d \xi, \nn
G_2 &=& \tfrac{1}{\sqrt{2}} s^{2/3} F_{(2)}, \nn
G_4 &=& - \sqrt{2} \tilde{g}^{-1} s^{1/3} c X^4 *_6 F_{(3)} \wedge d \xi - \tfrac{1}{\sqrt{2}} s^{4/3} X^{-2} *_6 F_{(2)} \nn
G_1 &=& - \tfrac{\sqrt{2}}{6} \tilde{g}^{-3} s^{1/3} c^3 \Delta^{-2} U d \xi -  \sqrt{2} \tilde{g}^{-3} s^{4/3} c^4 \Delta^{-2} X^{-3} d X, \nn
H^i_3 &=& \tfrac{1}{\sqrt{2}} g^{-2} s^{1/3} c \tilde{F}_{(2)}^i \wedge d \xi, \nn
H^i_{2} &=& - \tfrac{1}{2 \sqrt{2}} g^{-2} s^{4/3} c^2 \Delta^{-1} X^{-3} \tilde{F}^{i}_{(2)},
\eea
where
\bea
\label{deltaU}
\Delta &=& X c^2 + X^{-3} s^2 , \nn
U &=& X^{-6} s^2 - 3 X^2 c^2 + 4 X^{-2} c^2 - 6 X^{-2},
\eea
are given in terms of the scalar $X = e^{-\frac{1}{2 \sqrt{2}} \tilde{\phi}}$ and we have employed the shorthand $s \equiv \sin \xi, c \equiv \cos \xi$. Note also that $*_6$ denotes Hodge duality with respect to the six-dimensional space-time. Later, we will be interested in seven-dimensional Hodge duals, denoted $*_7$, and ten-dimensional Hodge duals which will appear without subscripts as in appendix \ref{sec:IIEOMs}.  Our conventions for Hodge duality follow \cite{Sfetsos:2010uq,Itsios:2012dc}
\be
\label{hodge}
(*_{D} F_p)_{\mu_{p+1} \dots \mu_{D}} = \frac{1}{p!} \sqrt{g} \e_{\mu_1 \dots \mu_D} F^{\mu_1 \dots \mu_{p}}_p,
\ee
where for ten-dimensions we take the sign $\e_{0 \dots 9} = +1$.

At this point it is also useful to record the orthonormal frame
\bea
\label{vielbein}
e^{\mu} &=& X^{-1/4} s^{-1/6} \Delta^{1/4} \bar{e}^{\mu}, \nn
e^{6} &=& \sqrt{2} \tilde{g}^{-1} X^{3/4} s^{-1/6} \Delta^{1/4} d \xi, \nn
e^{i} &=&  \tfrac{1}{\sqrt{2}} \tilde{g}^{-1} X^{-3/4} s^{-1/6} \Delta^{-1/4} c \,h^{i}.
\eea
We will employ this frame to perform checks on the derived equations of motion. In other words, we can take our equations of motion and plug in (\ref{connect}) and verify that one recovers the equations of motion of the theory (\ref{cveticaction}), which may be explicitly found in \cite{Cvetic:1999un}. We will see that the KK reduction from massive IIA on $S^4$ passes some non-trivial checks instilling confidence that it has been performed correctly.

%\textcolor{red}{It is useful to record the above fluxes in orthonormal frame:
%\bea
%G_1 &=& - \tfrac{4}{3} g^{-2} s^{1/3} c^3 \Delta^{-9/4} X^{-3/4} U e^6 - 8 \sqrt{2} g^{-3} s^{3/2} c^4 X^{-11/4} \Delta^{-9/4} \partial_{\mu} X e^{\mu}, \nn
%H_2^i &=& - \tfrac{1}{2 \sqrt{2}} g^{-2} s^{5/3} c^2 X^{-5/2} \Delta^{-3/2} \tfrac{1}{2!} (F^i_{(2)})_{\mu \nu} e^{\mu \nu}, \nn
%H_3^i &=& \tfrac{1}{2 } g^{-1} c s^{5/6} \Delta^{-3/4} X^{-1/4} \tfrac{1}{2!} (F^i_{(2)})_{\mu \nu} e^{\mu \nu 6}, \nn
%G_4 &=& - s c X^4 \Delta^{-1} \tfrac{1}{3!} (*_6 F_{(3)})_{\mu \nu \rho} e^{\mu \nu \rho 6} - \tfrac{1}{\sqrt{2}} s^2 X^{-1} \Delta^{-1} \tfrac{1}{4!} (*_6 F_{(2)})_{\mu \nu \rho \lambda} e^{\mu \nu \rho \lambda},
%\eea }

\subsection{Flux equations}
Observe that as we have only changed the four-form flux $F_4$, we simply have to ensure that all Bianchi identities and flux equations of motion involving $F_4$ are satisfied. We begin with the Bianchi identities.

The Bianchi identities for $H$ and $F_2$ are unchanged leading to $d H = 0$ and
\be \label{bianch0} d G_2 = m H.
\ee
In contrast, imposing the remaining Bianchi involving $F_4$ (\ref{BianchiIIA}) leads to
\bea
\label{bianch1} d G_4 - F^i \wedge H^i_3&=& H \wedge G_2, \\
\label{bianch2} H^i_{3} &=& G_1 \wedge F^i + d H^i_2 - \e_{ijk} H_2^j \wedge A^{k} \\
\label{bianch3} d G_1 &=& 0,
\eea
where we have defined $F^{i} = d A^i + \frac{1}{2} \e_{ijk} A^j \wedge A^{k}$. More concretely, (\ref{bianch1}) comes from expressions without $\sigma^i$, (\ref{bianch2}) comes from $\s^i \wedge \s^j$ terms and (\ref{bianch3}) comes from terms proportional to the volume of $S^3$. The terms proportional to $\sigma^i$ are simply the derivatives of (\ref{bianch2}). One can check that the equations here are consistent with the known reduction (\ref{connect})\footnote{To confirm this (11) of \cite{Cvetic:1999un} is useful.}.  This concludes discussion of the Bianchi identities.

Next we move onto the flux equations of motion (\ref{fluxIIAeq1}), (\ref{fluxIIAeq2}) and (\ref{fluxIIAeq3}), making use of the Hodge duals (\ref{IIAhodge}) as we go. We start with (\ref{fluxIIAeq2}) as the result is less involved. One encounters just two equations
\bea
\label{F2eq1} d ( e^{3 A} *_7 G_2) + e^{3A} H \wedge (*_7 G_4) &=& 0, \\
\label{F2eq2}  e^{2 A} (*_7 G_2) \wedge F^i -H \wedge (*_7 H_3^i) &=& 0.
\eea
As a consistency check one can confirm both of these against (\ref{connect}) and confirm that they are consistent with the reduction ansatz of Cveti\v{c} et al. \cite{Cvetic:1999un}.

From (\ref{fluxIIAeq3}), we get the following equations, which are respectively terms proportional to the volume of the $S^3$ , $\sigma^i \wedge \sigma^j$ and those without $\sigma^i$:
\bea
\label{F4eq1} d (e^{3 A} *_7 G_4 ) &=& -H \wedge G_1, \\
\label{F4eq2} d (e^{A} *_7 H_3^i) &=& \e_{ijk} e^{A} (*_7 H_3^j) \wedge A^k + e^{-A} *_7 H_2^i  \nn && \phantom{xxxxxxxxxxxxx}+ H \wedge H^i_2 + e^{3 A} *_7 G_4 \wedge F^i, \\
\label{F4eq3} d(e^{-3A} *_7 G_1) &=& - e^{-A} *_7 H_2^i \wedge F^i - H \wedge G_4.
\eea
Again one finds that the omitted equation is not independent and is simply the derivative of (\ref{F4eq2}) when one uses (\ref{bianch2}) and (\ref{F4eq1}). This is similar to what we noticed with the Bianchi, namely that the $\s^i$ conditions were implied. As a spot check of (\ref{F4eq3}) one can substitute (\ref{connect}) and using our conventions for the Hodge dual (\ref{hodge}), one recovers the last equation of (11) of \cite{Cvetic:1999un}.

%Some simplification of the last term is again required. So once all the terms are written out it is useful to substitute for the LHS of (\ref{F4eq3}). Some further massaging using (\ref{F4eq2}) leads finally to (\ref{F4eq4}).

Finally, we address the $B$-field equation of motion (\ref{fluxIIAeq1}). Decomposing this equation of motion we get the following two equations:
\bea
 d ( e^{-2 \Phi +3 A} *_7 H) &=&  e^{3 A} G_2 \wedge (*_7 G_4)  + G_4 \wedge G_1 - H^i_{3} \wedge H^i_{2} \nn && \label{Heq1} \phantom{xxxxxxxxxxxxxxx} + m e^{3A} *_7 G_2, \\
\label{Heq2} e^{-2 \Phi +3 A} *_7 H \wedge F^i &=& e^{A} G_2 \wedge (*_7 H_3^i) - G_4 \wedge H_2^i + \tfrac{1}{2} \e_{ijk} H_3^{j} \wedge H_3^{k}.
%\label{Heq3} \tfrac{1}{8} e^{-2 \Phi +3 A} *_7 H \wedge d F^i &=& 2 e^{-A} G_2 \wedge (*_7 H^i_2) + \e_{ijk} \tfrac{1}{2} e^{A} G_2 \wedge (*_7 H_3^j) \wedge A^k - G_4 \wedge H^i_3 \nn &-& \e_{ijk} G_4 \wedge H_2^j \wedge A^k + H_3^i \wedge \sum_j A^j \wedge H_3^j
\eea
Once more there is an extra equation, but after some massaging involving (\ref{bianch1}), (\ref{bianch2}),  (\ref{F2eq2}) and (\ref{Heq1}), one can show that this equation is simply the derivative of (\ref{Heq2}), so we can ignore it.

The above equations constitute all the flux equations of motion for our KK ansatz and lead to $D=7$ equations of motion. As the reader can observe, amongst these equations we also have various constraints such as (\ref{F2eq2}) and (\ref{Heq2}) which it may be difficult to imagine as arising from the process of varying an action. Indeed, we envisage that a more general KK ansatz will lead to a completion of some of these equations, so here we do not attempt to reconstruct the Lagrangian.

\subsection{Einstein \& dilaton equations}
\label{sec:EinsteinIIA}
In this subsection we work out the equations of motion which require a knowledge of the curvature. Choosing the natural orthonormal frame
\be
\label{frame1}
e^{\mu} = \bar{e}^{\mu}, \quad e^{i} = e^{A} ( \s^i - A^i),
\ee
where $\mu = 0, \dots, 6 $ and $ i =1, 2, 3$, using the spin connection (\ref{spin1}) one can determine the Ricci tensor
\bea
\label{ricci1} R_{11} &=& \tfrac{1}{2} e^{-2 A} - \nabla_{\rho} \nabla^{\rho} A - 3 \partial_{\rho} A \partial^{\rho} A  + \tfrac{1}{4} e^{2A} F^1_{\rho \mu} F^{1 \rho \mu}, \\
\label{ricci2} R_{\mu 1} &=& \tfrac{1}{2} e^{-4 A} D_{\rho} \left( e^{5A} F^{1\rho}_{~~\mu} \right), \\
\label{ricci3} R_{\mu \nu} &=& \bar{R}_{\mu \nu} - 3 \left( \nabla_{\nu} \nabla_{\mu} A + \partial_{\mu} A \partial_{\nu} A \right) - \tfrac{1}{2} e^{2A} F^i_{\mu \rho} F^{i~\rho}_{\nu}.
\eea
For simplicity we will just focus on a particular value for the $SU(2)$ index with the others following through a change of index. Here we have defined $D \omega^i = d \omega^i +  \e_{ijk} A^j \wedge w^k$ as in \cite{Cvetic:1999un}.

The Einstein equation is then
\bea
\label{Einstein1} R_{11} + 2 \partial^{\mu} A \partial_{\mu} \Phi&=& e^{2 \Phi} \biggl[ \tfrac{1}{4} e^{-6A} G_1^2 - \tfrac{1}{4}\left( \tfrac{1}{2} G_2^2 +\tfrac{1}{4!} G_4^2 +m^2 \right) \nn&+& \tfrac{1}{3!} e^{-2A} \left[ (H_3^1)^2 - (H_3^2)^2 - (H_3^2)^2 \right] \nn
&+& 2 e^{-4 A} \left[ (H_2^2)^2 + (H_2^3)^2 - (H_2^1)^2 \right] \biggr].
\eea
Observe that there is no $H$ along the internal $S^3$ so this drops out of (\ref{Einstein1}). It is also worth observing that since we get similar expressions for $R_{22}$ and $R_{33}$, the expected symmetry in the index $i$ implies the relationship
\be
\label{FH2H3}
\tfrac{1}{2!} e^{2A} (F^i)^2 =  e^{2 \Phi -2A} \left[ \tfrac{1}{3!} (H_3^i)^2- \tfrac{1}{2!} e^{-2A} (H_2^i)^2 \right].
\ee
Indeed, one can check that this is consistent with \cite{Cvetic:1999un}.

So we can write the Einstein equation along the $S^3$ in the following way
\bea
\label{ES3}
&& \tfrac{1}{2} e^{-2 A} - \nabla_{\rho} \nabla^{\rho} A - 3 \partial_{\rho} A \partial^{\rho} A  + 2 \partial^{\mu} A \partial_{\mu} \Phi \nn
&& \phantom{xxxxxxxxxx}= e^{2 \Phi} \biggl[ \tfrac{1}{4} e^{-6A} G_1^2 - \tfrac{1}{4}\left( \tfrac{1}{2} G_2^2 +\tfrac{1}{4!} G_4^2 +m^2 \right) \nn&& \phantom{xxxxxxxxxx} + \tfrac{1}{4} \left( -  \tfrac{1}{3!} e^{-2A} (H_3^i)^2
+ \tfrac{1}{2!} e^{-4 A}  (H_2^i)^2 \right)  \biggr].
\eea

One can also check that (\ref{Einstein1}) gives the scalar equation of motion of Romans' theory. This is a non-trivial check that this equation is correct.

We can now move onto the Einstein equation for the cross-terms. This necessitates that we calculate $\nabla_{\mu} \nabla_{i} \Phi$, a sketch of which can be found in the appendix for the simpler case where we have a $U(1)$ truncation of the $SU(2)$. Combining all the necessary terms one arrives at the equation
\bea
\label{iibEcross}
D_{\rho} \left( e^{5A- 2 \Phi} F^{i \rho}_{~~\mu} \right) &=& \biggl[ -e^{-A} G_{1 \rho} H_{2\mu}^{i~\rho} + e^{3A} \tfrac{1}{3!} H^i_{3 \rho \sigma \lambda} G_{4 \mu}^{~~\rho \sigma \lambda} \nn &+&  e^{A} \e_{ijk} \tfrac{1}{2!} H^j_{2\rho \sigma} H_{3\mu}^{k \rho \sigma} \biggr].
\eea

Finally we work out the Einstein equation for $M_7$. This takes the form
%(\textcolor{red}{to correct factors})
\bea
&& \bar{R}_{\mu \nu} - 3 (\nabla_{\nu} \nabla_{\mu} A + \partial_{\mu} A \partial_{\nu} {A} )  - \tfrac{1}{2} e^{2A} F^{i}_{\mu \rho} F^{i~\rho}_{\nu} + 2 \nabla_{\mu} \nabla_{\nu} \Phi - \tfrac{1}{4} H^2_{\mu \nu} \nn
&&= e^{2 \Phi} \biggl[ \tfrac{1}{2} e^{-6A} (G_1^2)_{\mu \nu} + \tfrac{1}{2} (G_2^2)_{\mu \nu} + \tfrac{1}{12} (G_4^2)_{\mu \nu} +  \tfrac{1}{2} e^{-4 A} (H^{i\;2}_2)_{\mu \nu} + \tfrac{1}{4} e^{-2A} (H^{i\;2}_3)_{\mu \nu} \nn
&&- \tfrac{1}{4} g_{\mu \nu} \left( e^{-6A} G_1^2 + \tfrac{1}{2} G_2^2 + \tfrac{1}{24} G_4^2 + m^2 + \tfrac{1}{2} e^{-4A} (H_2^i)^2 + \tfrac{1}{3!} e^{-2 A} (H_3^i)^2 \right)
\biggr].
\eea
In deriving this equation one has to determine an expression for $\nabla_{\mu} \nabla_{\nu} \Phi$ which may have a non-trivial dependence on the $S^3$ when the gauging is taken into account.  A calculation reveals that all dependence on the $S^3$ through the Christoffel symbols drops out so that $\nabla_{\mu} \nabla_{\nu} \Phi$ only depends on the seven-dimensional metric.

We can finally now work out the scalar curvature and determine the dilaton equation in type IIA. Since this equation only involves the NS sector and not the RR fields, this presents a convincing test for the corresponding KK reduction ansatz from type IIB. In other words, after non-Abelian T-duality we should encounter the same dilaton equation. We will comment on this in due course. For the moment, we contract the above Ricci tensors (\ref{ricci1}) and (\ref{ricci3}) and deduce that the dilaton equation takes the form
\bea
\label{iiaD}
0 &= & \bar{R} + \tfrac{3}{2} e^{-2 A} - 6 \nabla^2 A - 12 (\partial A)^2 + 12 \partial A \cdot \partial \Phi \nn
&+& 4 \nabla^2 \Phi - 4 (\partial \Phi)^2 - \tfrac{1}{12} H^2 - \tfrac{1}{4} e^{2 A} F^i_{\mu \nu} F^{i\mu \nu}.
\eea

\section{Reduction from IIB}
\label{sec:IIBred}
In this section we perform the analogous reduction on the non-Abelian T-dual. Simply by gauging the $S^2$, we will show that one can reinstate the $SU(2)$ gauge fields in a consistent way throughout. So the approach is this. Starting from the residual $S^2$ of the non-Abelian T-dual we gauge the $S^2$ in the natural way (see for example \cite{Gauntlett:2007sm}). This determines the metric and the dilaton is unchanged from \cite{Sfetsos:2010uq,Itsios:2012dc} since it is not sensitive to the gauging. The $B$-field follows from closure of the field strength $H = dB$ and one can confirm the NS sector is correct by reproducing the dilaton equation of the IIA reduction (\ref{iiaD}). Finally, we use knowledge of the NS sector to piece together the RR fields in a fashion that recovers the equations of motion of section \ref{sec:IIAred}.

\subsection{NS sector}
\label{sec:NS}
Recall from \cite{Sfetsos:2010uq,Itsios:2012dc} that, in the absence of $SU(2)$ gauge fields, an $SU(2$) transformation on $S^3$ leads to an internal metric of the form
\be
ds^2_{\textrm{T-dual}} = e^{-2A} dr^2 + \frac{r^2 e^{2A}}{r^2 + e^{4A}} ds^2(S^2).
\ee
If one wants to further gauge this residual $SU(2)$ isometry, the natural ansatz to consider is presented in appendix \ref{sec:S2gauge}. Assuming one proceeds in this fashion, one can anticipate the required form of the $B$-field from a knowledge of the $B$-field prior to gauging, namely
\be
\tilde{B} = B -  \frac{r^3}{r^2 + e^{4A}} \vol(S^2)
\ee
where tildes have been employed to differentiate the T-dual $B$-field from the original massive IIA one and we have flipped a sign from the $B$-field presented in \cite{Sfetsos:2010uq,Itsios:2012dc}. This sign flip is important and depends on the whether one is using left-invariant or right-invariant forms to parametrise the $S^3$. To date, all examples of $SU(2)$ transformations have assumed right-invariant forms \cite{Sfetsos:2010uq,Itsios:2012dc}, however here that choice is dictated by the ansatz of \cite{Cvetic:1999un} where left-invariant forms appear.

Now, we replace derivatives with gauge-covariant derivatives $D \mu^i = d \mu^i - \e_{ijk} \m^j A^k$ and closure of the field strength $\tilde{H} = d \tilde{B}$ leads to
\bea
\label{H}
\tilde{H} &=& d \tilde{B}, \nn
&=& H - \left[ \frac{r^2(r^2 + 3 e^{4A} )}{(r^2 + e^{4A})^2} dr - \frac{4 r^3 e^{4A}}{(r^2 + e^{4A})^2} d A\right] \wedge \vol(\tilde{S}^2) \nn
&+& \frac{r e^{4 A} }{r^2 + e^{4 A}} D \mu^i \wedge F^i+\mu^i F^i \wedge dr,
\eea
where $H = d B$, $F^i = d A^i + \frac{1}{2} \e_{ijk} A^j \wedge A^k$ and we can define the gauged $S^2$ with unit radius through the constrained variables $\mu^i \mu^i =1$ as
\be
 \vol (\tilde{S}^2) = \tfrac{1}{2} \e_{ijk} \mu^i D \mu^j D \mu^k.
\ee
Further details can be found in appendix \ref{sec:S2gauge}.

Note, in the non-Abelian dual only the one-forms $dr, D \mu^i$ appear making this the only choice and it is particularly easy to see this when one truncates the $SU(2)$ gauge fields to the Cartan $U(1)$ gauge field. In other words, $F^i$ has to appear with the $SU(2)$ index contracted and wedged with one of these forms. The transformed dilaton $\tilde{\Phi}$ is unchanged from \cite{Sfetsos:2010uq,Itsios:2012dc}, so we now have determined the NS sector and simply need to determine the RR fluxes in the next section \ref{sec:IIBred}. In fact, using the prescription for the $SU(2)$ transformation outlined in \cite{Itsios:2013wd} it is possible to generate the NS sector using non-Abelian T-duality, a procedure which we reproduce in appendix \ref{sec:NAT}.

So we can summarise the NS sector for the IIB KK reduction ansatz
\bea
\label{iibmet}
ds^2 &=& ds^2(M_7) + e^{-2A} dr^2 + \frac{r^2 e^{2A}}{r^2 + e^{4 A}} D \mu^i D \mu^i,  \\
\label{iibB}
\tilde{B} &=& B -  \frac{r^3}{r^2 + e^{4A}} \tfrac{1}{2} \e_{ijk}  \mu^i D \mu^j \wedge D \mu^k + A^i \wedge d (r \mu^i)  \nn
 &+&  r \tfrac{1}{2} \e_{ijk} \mu^i A^j \wedge A^k, \\
\label{iibdil} e^{-2 \tilde{\Phi}} &=& e^{-2 \Phi} e^{2A} (r^2 + e^{4A}).
\eea

To gain confidence that we are on the right path, we are now in a position to show that the dilaton equation using this KK ansatz for the NS sector reproduces the expected dilaton equation (\ref{iiaD}). Making use of the later Ricci tensor terms in section \ref{sec:EinsteinIIB}, the field strength (\ref{H}), the dilaton expression (\ref{iibdil}), in addition to the orthonormal frame
\be
D \mu^{i} = \frac{\sqrt{r^2 + e^{4A}}}{r e^{A}} (K^i_{\phi} e^1 - K^{i}_{\theta} e^2 ),
\ee
and appendix \ref{sec:S2gauge} where $K^i_{\theta}, K^i_{\phi}$ are defined,  a simple calculation is all that is required to reproduce (\ref{iiaD}) on the nose. This is a non-trivial check and a strong indication that the non-Abelian T-dual geometry can be gauged and reduced to give the same seven-dimensional theory.

\subsection{RR fluxes}
\label{sec:IIBredRR}
In this subsection we will infer the rest of the KK reduction ansatz since, as we have witnessed in the last subsection, we can now have full confidence in the NS sector. Recall that we inherit the mass $m$, fluxes $G_1, G_2$ and $G_4$ from \cite{Itsios:2012dc}, so we simply have to find the correct place for the fields $H_2^i$ and $H_3^i$ to enter. One subtlety is that as we started with left-invariant forms and not the usual right ones, even when $A^i = H_2^i = H_3^i$, we will not recover exactly the reduction ansatz of \cite{Itsios:2012dc}, but one with some signs flipped. We have identified which signs to change by resorting to our knowledge of non-Abelian T-duality, where the change in $SU(2)$ factor results in a flip in relative sign in the Lorentz transformation matrix $\Omega$ which acts on the spinors \cite{Sfetsos:2010uq,Itsios:2012dc}.

While the RR fluxes can be generated via non-Abelian T-duality (we sketch this calculation in appendix \ref{sec:NAT}), since we have to check the equations of motion regardless, here we opt to use  information about the NS sector KK reduction ansatz to piece together the missing parts. We begin with the one-form flux. Closure of this term, i.e. satisfying the Bianchi (\ref{BianchiIIB}), suggests strongly that this term does not change, modulo the sign flip imposed by the change of $SU(2)$ factor. This leads to
\be
\label{RRF1}
F_1 = - G_1+ m r dr.
\ee

% I have not got the non-Abelian transformation for the RR fluxes to work out yet, so we will work by trial and error to get the correct ansatz. In each case we should be recovering the equations of motion from section \ref{sec:IIAred}. It is hard to imagine the one-form flux changing as one would have to add a closed one-form to it. This cannot be built out of $A^i, H_2^i$ or $H_3^i$, so it seems natural that this will be unchanged from that of \cite{Itsios:2012dc}
%\be
%F_1 = - G_1 \textcolor{red}{+} m r dr.
%\ee

We now move onto the three-form flux and consider the following form, again with some sign changes to account for the change in $SU(2)$ factor,
\bea
\label{RRF3}
F_3 &=& e^{3A} *_7 G_4 + r dr \wedge G_2 + \frac{r^2}{r^2 + e^{4 A}} \left[ r G_1 + m e^{4 A} dr \right] \wedge \vol (\tilde{S}^2)  \nn
&-& r \mu^i H^i_3 - (r D \mu^i + \mu^i dr) \wedge H_2^i.
\eea
As an initial test of consistency, one can confirm that (up to signs) we recover the three-form presented in \cite{Itsios:2012dc} when we set the fields $A^i, H_2^i, H_3^i$ to zero. Essentially  the original field content can be found in the upper line and the lower line is constructed so that (\ref{bianch2}) is reproduced from the Bianchi identity (\ref{BianchiIIB}), $d F_3 = \tilde{H} \wedge F_1$, where $\tilde{H}$ can be found in (\ref{H}).  In addition, the Bianchi leads to the equations (\ref{bianch0}), (\ref{bianch3}) and (\ref{F4eq1}). Interestingly, even though our ansatz changes when we decide to do an $SU(2)$ transformation on a different $SU(2)$ factor, certain equations of motion such as (\ref{bianch0}) and (\ref{F4eq1}) do not change, meaning the the sign changes we have imposed have the correct structure. This is expected as we have used non-Abelian T-duality to confirm the required sign changes.

Now that we have discussed the one-form flux and found a three-form flux that reproduces some of the equations of motion exactly, it makes sense now to check this is consistent with (\ref{fluxIIBeq2}) since this is the remaining equation that couples these two flux terms. The respective Hodge duals are recorded in the appendix (\ref{IIBhodge}) and plugging these into the equation of motion we get the equations (\ref{F2eq2}) and (\ref{F4eq3}).

In deriving these expressions, it is useful to employ relationships such as
\bea
\left( \mu_2^2 + \mu_3^2 \right) \vol(\tilde{S}^2) &=& \mu_3 D \mu_1 \wedge D \mu_2 + \mu_2 D \mu_3 \wedge D \mu_1, \nn
\mu_1 \mu_2 \vol(\tilde{S}^2) &=& \mu^1 D \mu^3 \wedge D \mu^1,
\eea
and related cyclic expressions.

Finally, we come to the self-dual five-form flux. We start by changing the appropriate signs to account for the change in $SU(2)$ factor and then one can write down the correct ansatz using just a knowledge of the three-form, the $B$-field and the Bianchi identity for $F_5$. This determines the third line in the following expression by ensuring that terms proportional to derivatives of the warp factor $A$ vanish and the terms in the second line follow largely from the required self-duality of the five-form flux:
\bea
\label{RRF5}
F_5 &=& \frac{r^2 e^{3A}}{r^2 + e^{4 A}} \left( - r *_7 G_4 + e^{A} dr \wedge G_2 \right) \wedge \vol(\tilde{S}^2) - e^{3A} *_7 G_2 + r dr \wedge G_4 \nn
&-&  (r D \mu^i + \mu^i dr) \wedge e^A *_7 H_3^i - r \mu^i e^{-A} *_7 H_2^i - r H_3^i \wedge dr \wedge \e_{ijk} \mu^j D \mu^k \nn &+& \frac{r^3}{r^2 + e^{4 A}} \mu^i H_2^i \wedge dr \wedge \vol(\tilde{S}^2) - \mu^i \frac{r^2 e^{4A}}{(r^2 + e^{4A})} H_3^i \wedge \vol(\tilde{S}^2).
\eea

In addition to those identified earlier, the Bianchi identity for $F_5$ then leads to the following equations: (\ref{bianch1}), (\ref{F2eq1}) and (\ref{F4eq2}).  In deriving these equations, the following identities and their cyclic forms are useful
\bea
D \mu^i \wedge \vol(\tilde{S}^2) &=& 0, \nn
d (\mu^2 D \mu^3 - \mu^3 D \mu^2) &=& 2 \mu^1 \vol(\tilde{S}^2) - \mu^1 \sum_{i} \mu^i F^i + F^1 \nn
&-& (\mu^2 D \mu^1 - \mu^1 D \mu^2) \wedge A^2 - (\mu^3 D \mu^1 - \mu^1 D \mu^3) \wedge A^3.\nonumber
\eea
Last but not least, one can confirm that the remaining RR flux equation of motion (\ref{fluxIIBeq3}) offers nothing new and reproduces the equations we have identified above.

We now have expressions for all the RR fluxes and have determined our KK reduction ansatz from type IIB. Despite this, we still need to check the remaining equations of motion, namely the $B$-field equation of motion (\ref{fluxIIBeq1}) and the Einstein equation (\ref{EinsteinIIB}). We begin here with the $B$-field and in the next subsection we discuss the Einstein equation to show that the reduction is consistent. Plugging in our new $B$-field (\ref{iibB}), one recovers the two equations (\ref{Heq1}) and (\ref{Heq2}), and as is common for T-duality where one has mixing between cross-terms in the metric and $B$-fields, one is unsurprised to find the Einstein equation cropping up. Making use of $\mu^i D \mu^i =0$ and the relationship
\be
\label{F2H2H3}
e^{2A} \tfrac{1}{2!} F^i_{\mu \nu} F^{j\mu \nu} = e^{2 \Phi-2A} \left[ \tfrac{1}{3!} H^i_{3 \mu \nu \rho} H_3^{j\mu \nu \rho} - \tfrac{1}{2!} e^{-2A} H^i_{2\mu \nu} H_2^{j\mu \nu} \right],
\ee
which one can check is consistent with the reduction of Cveti\v{c} et al. using (\ref{connect}), one recovers the Einstein equation along $S^3$ (\ref{ES3}) and the equation corresponding to cross-terms in the metric (\ref{iibEcross}). Observe also that (\ref{F2H2H3}) is simply a generalised version of (\ref{FH2H3}).

\subsection{Einstein equation}
\label{sec:EinsteinIIB}
At this stage we have checked the dilaton equation and flux equations and found perfect agreement with the equations of motion resulting from the massive IIA reduction on the gauged $S^3$ presented in section \ref{sec:IIAred}. Therefore, it would be most surprising if the Einstein equations did not also conform. To check these we introduce a natural orthonormal frame for the metric (\ref{iibmet})
\bea
\label{frameiib}
e^{\mu} &=& \bar{e}^{\mu}, \nn
e^{r} &=& e^{-A} dr, \nn
e^{1} &=& \frac{r e^{A}}{\sqrt{r^2 + e^{4A}}} ( d \theta + \cos \phi A^1 - \sin \phi A^2), \\
e^{2} &=& \frac{r e^{A}}{\sqrt{r^2 + e^{4A}}} (\sin \theta d \phi - \cos \theta \sin \phi A^1 - \cos \theta \cos \phi A^2 - \sin \theta A^3 ). \nonumber
\eea
Using the derivatives (\ref{derivatives}) and the spin-connection (\ref{spiniib}) reproduced in the appendix, one can then calculate the Ricci tensor
\bea
\label{IIBricci1} R_{rr} &=& \nabla_{\rho} \nabla^{\rho} A + \frac{(r^2 - 3 e^{4A})}{(r^2 + e^{4 A})} (\partial A)^2+ \frac{6 e^{6 A}}{(r^2 + e^{4A})^2}, \\
\label{IIBricci2} R_{aa} &=& - \frac{(r^2 - e^{4A})}{(r^2 + e^{4A})} \nabla_{\rho} \nabla^{\rho} A - \frac{(r^4 - 12 r^2 e^{4 A} + 3 e^{8A}) }{(r^2 + e^{4A})^2} (\partial A)^2\nn
&+& \frac{(r^4 + 3 r^2 e^{4 A} + 6 e^{8 A})}{e^{2 A} (r^2 + e^{4A})^2} + \frac{r^2 e^{2 A}}{4(r^2 + e^{4 A})} K^i_{a} F^i_{\mu \rho} K^j_{a} F^{j \mu \rho}, \\
\label{IIBricci3} R_{\mu \nu} &=& \bar{R}_{\mu \nu}  - \frac{(r^2 - 3 e^{4 A}) }{(r^2 + e^{4 A} )} \nabla_{\mu} \nabla_{\nu} A - \frac{3 (r^4  -18 r^2 e^{4A} + e^{8A} )}{(r^2 + e^{4A})^2} \\
&-& \frac{r^2 e^{2 A}}{2(r^2 + e^{4A})} K^i_a F^i_{\mu \rho} K^j_{a} F^{j~\rho}_{~\nu}, \nn
\label{IIBricci4} R_{12} &=& \frac{1}{4} \frac{r^2 e^{2A}}{(r^2 + e^{4A})} K^i_{\theta} F^i_{\rho \sigma} K^j_{\phi} F^{j\,\rho \sigma}, \\
R_{ra} &=& 0, \\
\label{IIBricci5} R_{r \mu} &=& - \frac{12 r e^{5A} }{(r^2 + e^{4A})^2}, 
\eea
\bea
\label{IIBricci6} R_{a \mu} &=& \frac{r e^{A}}{2 \sqrt{r^2 + e^{4A}}} \biggl[ - K^i_{a}   \nabla_{\rho} F^{i~\rho}_{\mu} + K^i_{a}   \frac{(5 e^{4A} - 3 r^2)}{(r^2 + e^{4A})}  F^{i~\rho}_{\mu}\partial_{\rho} A \nn
&+& \e^{ab} K^{i}_{b} F^{i~\rho}_{\mu} \frac{1}{\sin \theta} \left( \sin \phi A^1_{\rho} + \cos \phi A^2_{\rho} \right) \biggr]
\eea
where we have introduced $a=1, 2$ (respectively $\theta$, $\phi$ directions) and the repeated index on the RHS of (\ref{IIBricci3}) is summed, whereas the indices in (\ref{IIBricci2}) are not.

We now comment on the Einstein equations and confirm that they also get mapped as expected. From both the diagonal $E_{rr}$ and $E_{aa}$ components of the Einstein equation we recover the Einstein equation along $S^3$ (\ref{ES3}). To make this connection we find that we have to use (\ref{F2H2H3}) and that the respective Einstein equations are related through the relationship
\be
E_{aa} = - \frac{(r^2 - e^{4A})}{(r^2 + e^{4A})} E_{rr}.
\ee

Moving on, one can check that the $E_{ra}$ component of the Einstein equation is satisfied. In contrast to the situation presented in \cite{Itsios:2012dc} where the $S^2$ is not gauged, here a cancellation is required. While both the Ricci tensor $R_{ra}$ and the term $\nabla_{r} \nabla_{a} \tilde{\Phi}$ are zero, (\ref{F2H2H3}) is required so that the flux terms disappear. The $E_{12}$ component of the Einstein equation is also satisfied for similar reasons, but here $R_{12}$ is not zero and has to combine with the contraction of the $\tilde{H}$ field strength in the correct fashion.

The $E_{r\mu}$ component of the Einstein equation, making use of (\ref{IIBricci5}),  is satisfied through various cancellations. In addition, one needs to make use of the identity
\be
\label{newrel}
F^i_{\rho \sigma} H_{\mu}^{~\rho \sigma} = e^{2 \Phi - 3A} \left[ H_{2\,\rho \sigma}^i (*_7 G_4)_{\mu}^{~\rho \sigma} + e^{A} G_{2\,\rho \sigma} H_{3\,\mu}^{i\,~\rho \sigma} \right].
\ee
One can check this is consistent with the KK reduction of \cite{Cvetic:1999un} by plugging in (\ref{connect}). Finally, a lengthier calculation reveals that various terms of the $E_{a \mu}$ Einstein equation conspire to reproduce (\ref{iibEcross}), where again one has to use (\ref{newrel}).

\subsection*{Summary}
In this section we have illustrated how the KK ansatz comprising of (\ref{iibmet}), (\ref{iibB}), (\ref{iibdil}) and the one-form (\ref{RRF1}), three-form (\ref{RRF3}) and five-form fluxes (\ref{RRF5}),  when plugged into the equations of motion of type IIB supergravity,  leads to the same equations of motion of the Cveti\v{c} et al KK reduction ansatz in $D=7$. More importantly, as we also check that non-Abelian T-duality leads to the same result in appendix \ref{sec:NAT}, we can confirm that non-Abelian T-duality is a symmetry of the equations of motion for a reasonably general ansatz. 

From $D=7$ using (\ref{connect}) we can further reduce to $D=6$ to recover the equations of motion of Romans' theory. So, we can safely conclude that any solution to Romans' F(4) gauged supergravity can be uplifted to type IIB supergravity using our KK reduction ansatz.

\section{Uplifted Solutions}
\label{sec:solns}

Having identified a consistent reduction from type IIB supergravity to Romans' F(4) gauged supergravity, in this section we generate some examples of new type IIB solutions. We start by considering examples with supersymmetry, notably a domain wall \cite{Lu:1995hm} and the ``magnetovac" identified originally by Romans \cite{Romans:1985tw}, which also serves as one end-point of the supersymmetric flows discussed in \cite{Nunez:2001pt}. While the former does not excite $SU(2)$ gauge fields, its inclusion here is motivated by the fact that it is an example of a supersymmetric geometry with a non-trivial scalar and may be regarded as an immediate generalisation of the supersymmetric $AdS_6$ vacuum, where the scalar is constant. Later in this section, we present the uplift of a geometry that fits into the class of Lifshitz geometries \cite{Kachru:2008yh}, which is itself a non-supersymmetric deformation of the magnetovac, before presenting a simple charged black hole first presented in \cite{Cvetic:1999un}, but here in its alternative type IIB setting.

Recall that the striking result of \cite{Lozano:2012au} was that one had the freedom to perform a non-Abelian T-duality on the warped $AdS_6 \times S^4$ solution of massive IIA to generate a solution of type IIB. From the lower-dimensional perspective, this discovery means that starting from the $AdS_6$ vacuum, we can either uplift to massive IIA or type IIB and supersymmetry remains unaffected. Since we are working in the context of ten-dimensional type II supergravity and the $AdS_6$ vacua require the presence of a geometric $SU(2)$ R-symmetry, it could be expected that the supersymmetric structures of both uplifts are the same. Through studying the uplifts of supersymmetric solutions in subsection \ref{sec:domain} and \ref{sec:magnetovac} we will produce evidence to support this claim. Naturally, the reduction of the Killing spinor equations would help to confirm our suspicions, but such an act falls outside of the scope of this work and we leave it to future work.

\subsection{Supersymmetric domain wall}
\label{sec:domain}
In addition to non-supersymmetric domain walls interpolating between the supersymmetric and non-supersymmetric $AdS_6$ vacua of F(4) gauged supergravity \cite{Gursoy:2002tx,Karndumri:2012vh}, supersymmetric domain walls also exist \cite{Lu:1995hm}.  Though the solution does not excite the $SU(2)$ gauge fields and is supported solely through the scalar field, it provides a less-trivial example of a supersymmetric solution.

Taking into account  a flip in metric signature from the conventions of Romans and an appropriate rescaling of the scalar $\phi$, the solution \cite{Lu:1995hm} reads
\bea
ds^2_6 &=& e^{2 B} \eta_{\mu \nu} dx^{\mu} dx^{\nu}+ e^{6 B} du^2, \nn
\phi &=& \frac{1}{\sqrt{2}} \log u, \nn
e^{-3B} &=& \frac{3}{2 \sqrt{2}} m \, u^{-1/2} - \frac{1}{2 \sqrt{2}} g \, u^{3/2}.
\eea
Using the Killing spinor equations of Romans \cite{Romans:1985tw}, it is easy to check that this domain wall solution preserves half the original supersymmetry and that the Killing spinors $\e_i$ satisfy
\be
\label{KS}
\e_i = e^{\frac{1}{2} B} \e^0_i, \quad \g_{u} \g_{7} \e_i^0 = \e_i^0,
\ee
where $\e_i^0$ denotes a constant spinor and $i$ is a $USp(4)$ vector index.

We now would like to uplift this solution to ten-dimensions. Since our interest here is supersymmetry, and in particular how it survives the uplifting process, it is instructive to first uplift the solution to massive IIA supergravity using \cite{Cvetic:1999un}, before later repeating the process to get a type IIB solution. As we will observe, despite the ease at which one can identify supersymmetries in the lower-dimensional theory, here for the uplifted solution the task becomes a lot less tractable, suggesting that the Killing spinors of Romans' theory (\ref{KS}) are related to those of massive IIA in a rather complicated fashion. So, for simplicity, we will make a particular choice for $g$ and $m$ by adopting
\be
g = 3m = 2 \sqrt{2}.
\ee
En route to performing the initial uplift to IIA,  we take the opportunity to identify various fields which are common to both IIA and IIB KK reduction ans\"{a}tze through (\ref{connect}):
\bea
\label{domainquants}
X &=& u^{1/2}, \nn
\Delta &=& u^{1/2} \tilde{\Delta} = u^{1/2} \left[ c^2 + u^{-2} s^2 \right], \nn
U &=& u^{-3} s^2 - 3 u c^2 + 4 u^{-1} c^2 - 6 u^{-1}, \nn
e^{A} &=& \frac{ \tilde{\Delta}^{-1/4} s^{-1/6} c }{2 u^{1/2}}, \nn
G_1 &=& - \frac{1}{12} s^{1/3} c^3 u^{-1} \tilde{\Delta}^{-2} U d \xi - \frac{1}{4} s^{4/3} c^4 \tilde{\Delta}^{-2} u^{-3} du.
\eea
Proceeding, following \cite{Cvetic:1999un} and employing the rewriting (\ref{cveticconn}), one arrives at the uplifted solution in massive IIA
\bea
\label{domainIIA}
ds^2_{10} &=&  s^{-1/3} \tilde{\Delta}^{1/2} \left[ ds^2_6 + u d \xi^2     + \frac{1}{4 u} \tilde{\Delta}^{-1} c^2 (\s^i)^2    \right], \nn
F_4 &=& - \left[ \frac{1}{12} s^{1/3} c^3 u^{-1} \tilde{\Delta}^{-2} U d \xi + \frac{1}{4} s^{4/3} c^4 u^{-3} \tilde{\Delta}^{-2} du \right] \s^1 \wedge \s^2 \wedge \s^3, \nn
e^{\Phi} &=& s^{-5/6} \tilde{\Delta}^{1/4} u^{-1/2},
\eea
and one can check that this is indeed a solution, thus again confirming that the ansatz provided in \cite{Cvetic:1999un} does what it claims to do.  In checking the equations, it should be borne in mind that the mass parameter of massive IIA is related to the gauge coupling \cite{Cvetic:1999un} through the relationship
\be
\label{IIAm}
\tilde{m} = \frac{\sqrt{2} }{3} \tilde{g} = \frac{(3 m g^3)^{1/4}}{3 \sqrt{2}},
\ee
where $m, g$ are now the original parameters in Romans' theory. Throughout this section we will use $\tilde{m}$ to denote the mass parameter of massive IIA supergavity on the understanding that it is not independent and is related to the gauge coupling of \cite{Cvetic:1999un} through (\ref{IIAm}).

Since the lower-dimensional solution breaks half the supersymmetry of the $AdS_6$ vacuum and we are also assuming that supersymmetry is preserved in the uplift to IIA, we anticipate that the solution (\ref{domainIIA}) preserves eight supersymmetries. To test this claim we evaluate the dilatino variation, which takes the form\footnote{We follow the supersymmetry conventions of \cite{hassan} and use the explicit gamma matrices in the appendix of \cite{Bakhmatov:2011aa}.}
\bea
\delta \lambda &=& M \eta, \nn
&=& \biggl[ \frac{1}{12} \left( -5 c + 2 (1-u^{-2}) s^2 c \right) \tilde{\Delta}^{-3/2} \G^6 + \frac{1}{2}\left(-\frac{s}{u } + \frac{s c^2}{2 u \tilde{\Delta}} \right) \tilde{\Delta}^{-1/2} (1-u^2)   \G^5 \nn &+& \frac{5}{12} \s^1 -\frac{1}{12} s \tilde{\Delta}^{-3/2} U \G^{6789} \s^1 - \frac{1}{4} c \,s^2 \tilde{\Delta}^{-3/2} (u^{-2} -1) \G^{5789} \s^1 \biggr] \eta.
\eea
Owing to the inherent complexity of the dilatino variation, explicitly showing supersymmetry and extracting the projection conditions would appear to be a difficult task. Instead, as supersymmetry is expected, we may check that the determinant of $M$ is zero, which implies that zero is an eigenvalue, i.e. there is some unbroken supersymmetry. Furthermore, one can show that there are eight zero eigenvalues corresponding to the eight expected supersymmetries. While, we have not solved the Killing spinor equations of massive IIA, and do not claim that we have, through looking at the dilatino variation we have observed that it is consistent with our expectation that eight supersymmetries are preserved.

We now move onto the non-Abelian dual and the uplift to type IIB. Taking note of the above expressions (\ref{domainquants}), the uplifted string frame IIB solution is
\bea
ds^2_{10} &=& s^{-1/3} \tilde{\Delta}^{1/2} \left[ ds^2_6 +  u^2  d \xi^2 \right] + e^{-2A} dr^2 + \frac{r^2 e^{2A}}{r^2 + e^{4A}} ds^2(S^2), \nn
B &=& -\frac{r^3}{r^2 + e^{4A}} \vol(S^2), \quad e^{\Phi} = \frac{\tilde{\Delta}^{1/4}}{s^{5/6} u^{1/2} e^{A} \sqrt{r^2 + e^{4A}}}, \nn
F_1 &=& - G_1 + \tilde{m} r dr, \nn
F_3 &=& \frac{r^2}{r^2 + e^{4A}} \left[r G_1 + \tilde{m} e^{4A} dr \right] \wedge \vol(S^2).
\eea
This bears a strong resemblance to (11) of \cite{Lozano:2012au}, but on closer inspection, one will see that $G_1$ and $e^{A}$ now have a dependence on the coordinate $u$.

We can now check supersymmetry of the non-Abelian T-dual relatively quickly. From earlier work \cite{Itsios:2012dc, Lozano:2012au} it is known that in the absence of the $SU(2)$ gauge fields, which is the case here, that the additional Killing spinor equations of the non-Abelian T-dual can be whittled down to a single expression
\be
\label{singlecond}
\left[ \frac{1}{2} \slashed{\partial} A \G_r - \frac{e^{-A}}{4} \G^{\a_1 \a_2} \s^3 - \frac{e^{\Phi}}{8} \left( \tilde{m} i \s^2 + e^{-3A} \slashed{G}_1 \G^{r \a_1 \a_2} \s^1 \right) \right] \eta = 0,
\ee
where $\a_i$ refer to directions on the two-sphere. Note here again that the change in the $SU(2)$ factor utilised in T-duality leads to a change in some signs. As explained in \cite{Itsios:2012dc}, the non-Abelian T-dual will now preserve the eight supersymmetries of the original geometry provided this condition breaks no further supersymmetries. So one has to make sure that the supersymmetries corresponding to zero eigenvalues of the above matrix agree with the eight Killing spinors of the original background. One finds that (\ref{singlecond}) preserves sixteen Killing spinors, eight of which can be mapped to the preserved supersymmetries of the original massive IIA solution. As such, the background preserves eight supersymmetries and we see that non-Abelian T-duality preserves the supersymmetry of the original domain wall solution. So we have seen that even with a non-trivial scalar profile that supersymmetry is preserved in the uplifts. In the next subsection we turn on a $U(1)$ gauge field.

\subsection{Supersymmetric magnetovac}
\label{sec:magnetovac}
\label{sec:ads4h2}
One of the simplest supersymmetric solutions to Romans' theory with $SU(2)$ gauge fields excited was identified by Romans in his original paper \cite{Romans:1985tw} and corresponds to the direct product $AdS_4 \times H^2$ where the field strength supporting the geometry is purely magnetic leading to a so-called ``magnetovac" solution. This solution also appeared as a fixed-point in the supersymmetric flows identified in \cite{Nunez:2001pt} and forms the basis of the Lifshitz solutions presented in \cite{Gregory:2010gx}, since the latter may be regarded as deformations of the $AdS_4$ space-time with dynamical exponent $z$. As the relativistic $AdS_4$ solution is recovered when $z=1$, these solutions are intimately related and we will discuss the Lifshitz solution in the next subsection.

We begin by identifying the original supersymmetric $AdS_4 \times H^2$ solution of Romans' theory and its massive IIA supergravity uplift. In the original notation of Romans \cite{Romans:1985tw} the solution may be expressed as
\bea
ds^2_6 &=&  \frac{1}{m^{2}} \left[  \frac{2 (dt^2 - dx_i^2 - dr^2)}{r^2} - \frac{dx^2 + dy^2}{y^2} \right], \nn
F^3_{(2)} &=& \frac{1}{2 m } \frac{dx \wedge dy}{y^2} , \quad \phi = 0,
\eea
where the signature of the metric follows from the mainly minus signature employed by Romans \cite{Romans:1985tw} and $(x,y)$ parametrise the hyperbolic space $H^2$. In addition, we have employed a global symmetry of Romans' theory to set the scalar to zero. This in turn means that gauge coupling $g$ and the mass $m$ are then related through $g = 2m$. If one chooses not to rescale $\phi$ to zero, more generally one finds the analysis in \cite{Nunez:2001pt} where $m$ and $g$ are independent\footnote{In \cite{Nunez:2001pt} the parameter $a$ in (25) is not free and for the Einstein equation to be satisfied for the solution presented here we require $a^{-1} = 2 m$.}.

To perform the uplift from Romans' theory one again has to employ (\ref{cveticconn}) to bring it to a form consistent with  \cite{Cvetic:1999un}.
In the notation of \cite{Cvetic:1999un} we now have
\be
\label{X}
X = e^{- \frac{1}{2 \sqrt{2}} \tilde{\phi}} = \left( \frac{2 }{3}\right)^{\frac{1}{4}},
\ee
where we have used $g = 2m$.

For the purposes of the uplift it would certainly simplify expressions if one could set $X=1$ by choosing a different constant for the scalar $\phi$ of Romans' theory. Indeed, Romans originally chooses $\phi =0$, but we know from the work of \cite{Nunez:2001pt} that more generally we have
\be
e^{2 \sqrt{2} \phi} = \frac{2m}{g}
\ee
at the supersymmetric fixed-point. A short calculation then shows that $g$ and $m$ generically drop out and $X$ always takes the value (\ref{X}). Therefore, no matter what form we take for the $AdS_4 \times H^2$ solution of Romans' theory, the uplift will involve unsightly factors of $X$ being retained.

%So we could attempt to set $X=1$ but from the above relationships (\ref{cveticconn}) one finds that this would require ${m}/{g} = 0$.  Therefore, no matter what form we take for the $AdS_4 \times H^2$ solution of Romans' theory, the uplift will involve unsightly factors of $X$ being retained.

In addition to $X$, the following functions appear in the KK reduction ansatz
\bea
\tilde{g} &=& X^{-1}m, \nn
\Delta &=& \frac{1}{(2^3 3)^{\frac{1}{4}}} \left[ 2 + s^2 \right] =  \frac{1}{(2^3 3)^{\frac{1}{4}}} \tilde{\Delta}, \nn
U &=& \frac{\sqrt{3}}{2 \sqrt{2}} \left[ c^2 - 9 \right] = \frac{\sqrt{3}}{2 \sqrt{2}} \tilde{U}.
\eea

Putting everything together we determine the form for the IIA solution in string frame
\bea
\label{iiaAdS4H2}
ds^2_{10} &=& \frac{1}{\sqrt{2} m^2} s^{-1/3} \tilde{\Delta}^{{1}/{2}} \biggl[ 2 ds^2(AdS_4) + ds^2(H^2) + \frac{4}{3} d \xi^2    \nn
&+&  \tilde{\Delta}^{-1} c^2  \left( (\s^1)^2 + (\s^2)^2 + (\s^3 - \frac{dx}{y})^2 \right) \biggr] \nn
e^{\Phi} &=& {3}^{{1}/{4}} 2^{-{1}/{2}}  s^{-5/6} \tilde{\Delta}^{{1}/{4}}, \quad B = 0, \nn
F_2 &=& 0, \nn
F_4 &=& - m^{-3} 2^{{1}/{4}}  3^{-{3}/{4}} s^{1/3} c^3 \tilde{\Delta}^{-2} \tilde{U} d \xi \wedge h^3 \wedge \s^{12} \\
&+& m^{-3}  2^{-3/4} 3^{-3/4} \vol(H^2) \wedge ( 2 s^{1/3} c  h^3 \wedge d \xi - 3 s^{4/3} c^2 \tilde{\Delta}^{-1} \s^{12} ). \nonumber
\eea
Again when checking the equations of motion, it is good to recall (\ref{IIAm}).

As for supersymmetry, we again expect that supersymmetry is respected in the uplifting process. Here we confirm that the dilatino variation is consistent with unbroken supersymmetry. Plugging in the above solution into the dilatino variation one arrives at
\bea
&&\biggl[ - {c}  \frac{2^{3/2} [5 + s^2]}{5 \tilde{\Delta}^{3/2}} \G^{6} \s^1- \frac{\sqrt{3}}{5} \tilde{\Delta}^{-3/2} \tilde{U} s \G^{69 78} +  \frac{\sqrt{3}}{5} s \tilde{\Delta}^{-1/2} \G^{4596} \nn && \phantom{xxxxxxxxxxxx} - \frac{3}{5} s^2 \tilde{\Delta}^{-1} \G^{4578} + \mathbb{1}_{32} \biggr] \eta = 0. 
\eea
As noted in the previous subsection, the extraction of projection conditions from here looks involved, so we simply check that the determinant of the above matrix vanishes and that it supports eight zero eigenvalues corresponding to the expected eight supersymmetries. So, here again we recognise that a lower-dimensional supersymmetric solution when uplifted to massive IIA leads to a solution which is consistent with preserved supersymmetry.

We can now turn to the task of reading off a new $AdS_4 \times H^2$ solution to type IIB supergravity by determining the various components of the dual geometry. In terms of our notation, one identifies the following
\bea
e^{A} &=& 2^{-1/4} m^{-1} s^{-1/6} \tilde{\Delta}^{-1/4} c, \nn
A^3 &=& \frac{dx}{y}, \nn
G_1 &=& - m^{-3} 2^{1/4} 3^{-3/4} s^{1/3} c^3 \tilde{\Delta}^{-2} \tilde{U} d \xi.
\eea
Substituting these into our KK reduction ansatz from type IIB we find the full solution
\bea
\label{iibAdS4H2}
ds^2 &=& \frac{1}{\sqrt{2} m^2} s^{-1/3} \tilde{\Delta}^{{1}/{2}} \biggl[ 2 ds^2(AdS_4) + ds^2(H^2) + \frac{4}{3} d \xi^2 \biggr]   +
 e^{-2A} dr^2 \nn &+& \frac{r^2 e^{2A}}{r^2 + e^{4A}} \left[ d \theta^2 + \sin^2 \theta (d \phi - \frac{dx}{y})^2 \right]  \nn
e^{\Phi} &=& \frac{{3}^{{1}/{4}}   \tilde{ \Delta}^{{1}/{4}}}{ 2^{1/2} s^{5/6} e^{A} \sqrt{r^2 + e^{4A}}}, \nn
B &=& -  \frac{r^3}{r^2 + e^{4A}} \vol(\tilde{S}^2) - \frac{dx}{y} \wedge d (r \cos \theta)  \nn
F_1 &=& -G_1 + \tilde{m} r dr, \nn
F_3 &=& \frac{r^2}{r^2 + e^{4A}} \left[ r G_1 + \tilde{m} e^{4A} dr \right] \wedge \vol(\tilde{S}^2)  \\
&+& m^{-3} 2^{-3/4} 3^{-3/4} s^{1/3} c \vol(H^2) \wedge \left[ 2 r \cos \theta d \xi - 3 s c \tilde{\Delta}^{-1}  d( r \cos \theta) \right]  , \nn
F_5 &=& (1+*) \biggl[ r \,m^{-3} 2^{1/4} 3^{-3/4}  s^{1/3} c \sin^2 \theta \vol(H^2) \wedge d \xi \wedge dr \wedge \left(d \phi - \frac{dx}{y} \right)  \nn &+& \frac{r^2 \cos \theta s^{1/3} c}{2^{3/4} 3^{3/4} m^3 (r^2 + e^{4A})} \vol(H^2) \wedge \vol(\tilde{S}^2) \wedge \left( 3 r s c \tilde{\Delta}^{-1} dr + 2 e^{4A} d \xi\right) \biggr]. \nonumber
\eea

As before, we would now like to get some confirmation that supersymmetry is preserved. The expectation is that eight supersymmetries will survive the uplift to type IIB and an analysis of the dilatino variation of the geometry (\ref{iibAdS4H2}) reveals that the determinant of the dilatino variation vanishes and eight zero eigenvalues exist\footnote{The complexity of the solution meant that in performing this check we simply sampled the variation for particular values of the coordinates $(r, \xi, \theta)$.}, indicating that supersymmetry remains unbroken in the uplift to type IIB.

\subsection{Lifshitz}
Along with \cite{Balasubramanian:2010uk,Donos:2010tu,Donos:2010ax},  one of the earliest examples of string theory manifestations of geometries with Lifshitz symmetry \cite{Kachru:2008yh} was presented in \cite{Gregory:2010gx}. Setting it apart from direct constructions in higher-dimensions \cite{Balasubramanian:2010uk,Donos:2010tu,Donos:2010ax}, \cite{Gregory:2010gx} searched for Lifshitz configurations in lower-dimensional massive supergravities and isolated a particular class of solutions to Romans' theories both in five and six-dimensions. Here we review the six-dimensional solution, discuss the uplift to massive IIA and present an analogous solution to type IIB supergravity. As shown explicitly in \cite{Gregory:2010gx} these solutions are not supersymmetric, so stability is always going to be a concern, and, indeed, preliminary studies hint at the existence of instabilities \cite{Braviner:2011kz} whose physical significance has yet to be properly investigated.

But returning to the solution, in the notation of Romans (\ref{romansmassive}), the six-dimensional Lifshitz solution may be written as
\bea
ds^2_6 &=& L^2 \left[ r^{2z} dt^2 - r^2 (dx_1^2 + dx_2^2) - \frac{dr^2}{r^2}  - a^2 ds^2(H^2) \right], \nn
F^3 &=& e^{\phi_0 /\sqrt{2}} L \gamma \left[ \sqrt{z-1} \, r^{z-1} dt \wedge dr + a^2 \vol(H^2) \right] \nn
B &=& \tfrac{1}{2} e^{-\sqrt{2} \phi_0} L^2 \sqrt{z-1} \,r^2 dx_1 \wedge dx_2,
\eea
where for simplicity we have performed the rescalings of (2.17) and (2.18) of \cite{Gregory:2010gx} directly on the solution and dropped hats. Our un-hatted parameters are simply the hatted ones of \cite{Gregory:2010gx}. Above $z$ is the dynamical exponent, $\phi_0$ is a constant value of the Romans' scalar field, $\gamma, a$ are parameters we will define below, and $L$ is a scale corresponding to the $AdS_4$ radius when $z=1$. While the supersymmetric $AdS_4 \times H^2$ solution of section \ref{sec:ads4h2} is naturally recovered when $z=1$, more generally one can have $z \neq 1$ solutions where the parameters depend on the dynamical exponent \cite{Gregory:2010gx}
\bea
%\beta^2 &=& z-1, \nn
%\alpha^2 &=& \gamma^2 (z-1), \nn
\gamma^2 &=& \frac{(2+z)(z-3) \pm 2 \sqrt{2(z+4)}}{2z}, \nn
g^2 &=& 2 z (4 +z), \nn
\frac{m^2}{2} &=& \frac{6 + z \mp 2 \sqrt{2(z+4)}}{z}, \nn
a^{-2} &=& 6 + 3 z \mp 2 \sqrt{2 (z+4)}.
\eea
As explained in \cite{Gregory:2010gx}, this solution can be uplifted to massive IIA using the KK reduction ansatz of \cite{Cvetic:1999un}\footnote{In the uplifted solution presented in \cite{Gregory:2010gx} a notable typo concerns the RR two-form $F_2$ which cannot be zero, since otherwise the Bianchi identity is not satisfied. }. Alternatively, using our reduction ansatz the six-dimensional solution can be uplifted leading to a new solution of type IIB supergravity. The ten-dimensional metric exhibiting Lifshitz symmetry may be written as
\bea
ds^2 &=& X^{-1/2} s^{-1/3} \Delta^{1/2} \biggl[ -ds^2_6 + 2 \tilde{g}^{-2} X^2 d \xi^2 \biggr] + e^{-2A} dr^2 \\
&+&  \frac{r^2 e^{2A}}{r^2 + e^{4A}} \left( d \theta^2 + \sin^2 \theta \left( d \phi + e^{\phi_0 /\sqrt{2}} L \gamma \left[ z^{-1}\sqrt{z-1} \, r^{z} dt  - a^2 \frac{dx}{y} \right] \right)^2 \right),  \nonumber
\eea
where $X = e^{\phi_0/\sqrt{2}} (g/3m)^{1/4}$ and $e^{A}$ is defined in (\ref{connect}). We omit details of the rest of the solution but it can be pieced together from section \ref{sec:IIBred}.

\subsection{Black Holes}
To the extent of our knowledge, the most general black hole solution to Romans' theory was presented in \cite{Chow:2008ip}. The solution corresponds to a non-extremal charged rotating black hole with five parameters: a mass parameter $m$, two angular rotation parameters  $a, b$ describing motion in orthogonal two-planes, a single charge parameter $\delta$, and lastly the $SU(2)$ gauge coupling $g$. All of the charged solutions are supported solely through the excitation of a single $U(1)$ gauge field from the $SU(2)$ gauge group, so none of the charged black holes may be regarded as truly non-Abelian in nature, and as a direct consequence only the charge $\delta$ appears. Within this class of solutions one also finds supersymmetric solutions with expected zero temperature \cite{Chow:2008ip}.

This general solution  \cite{Chow:2008ip} threads together multiple strands of the literature and simpler solutions are recovered when various parameters are set to zero. For example, without charge, the solution reduces to the Kerr-AdS solution \cite{Hawking:1998kw,Gibbons:2004js, Chen:2006xh}, while minus the gauging, $g=0$, the solution corresponds to the Cveti\v{c}-Youm two-charge solution \cite{Cvetic:1996dt}. Finally, in the absence of rotation, $a =b = 0$, one finds the static solution of \cite{Cvetic:1999un} which, neglecting the supersymmetric $AdS_6$ vacuum \cite{Romans:1985tw, Brandhuber:1999np}, was the first solution to be uplifted to massive IIA using the KK reduction ansatz of \cite{Cvetic:1999un}. Given the parallels of our work to that of Cveti\v{c} et al., here we focus on the same solution and present an alternative uplift to IIB, though we point out that there is no obstacle to also uplifting the most general solution \cite{Chow:2008ip}.

In the notation of the action (\ref{cveticaction}), the six-dimensional solution takes the form\footnote{Here we take $k=1$ for simplicity.}
\bea
ds^2_6 &=& - H^{-3/2} f dt^2 + H^{1/2} \left( f^{-1} dr^2 + r^2 d \Omega_4^2 \right), \nn
\tilde{\phi} &=& \frac{1}{\sqrt{2}} \log H, \quad \tilde{A}^3_{(1)} = \sqrt{2} (1-H^{-1} )  \coth \beta dt, \nn
f &=& 1 - \frac{\mu}{r^3} + \frac{2}{9} g^2 r^2 H^2, \quad H = 1 + \frac{\mu \sinh^2 \beta}{r^3}.
\eea
To perform either the uplift to massive IIA or type IIB, one just needs to employ the ansatz of \cite{Cvetic:1999un} or our ansatz presented in section \ref{sec:IIBred} with $X = H^{-1/4}$. The string frame metric for the IIB solution takes the form
\bea
ds^2 &=& H^{1/8} s^{-1/3} \Delta^{1/2} \left[ ds^2_6 + 2 \tilde{g}^{-2} H^{-1/2} d \xi^2 \right]  + e^{-2A} dr^2 \\
&& \phantom{xxxxxxxxxx} + \frac{r^2 e^{2A}}{r^2 + e^{4A}} \left[ d \theta^2 + \sin^2 \theta \left(d \phi -  \sqrt{2} (1-H^{-1} )  \coth \beta dt \right)^2 \right].  \nonumber
\eea
The rest of the solution can be worked out using the expressions in section \ref{sec:IIBred}.

\section{Concluding Remarks}
In this work we have identified a recently discovered supersymmetric $AdS_6$ solution of type IIB supergravity \cite{Lozano:2012au} as the IIB uplift of the supersymmetric vacuum of Romans' F(4) gauged supergravity \cite{Romans:1985tw}. While this observation could have been made in the light of the results of \cite{Itsios:2012dc}, here we have completed the KK reduction ansatz to include the characteristic $SU(2)$ gauge fields and shown that this ansatz, via the type IIB equations of motion, leads to the equations of motion of Romans' theory. Therefore, any solution to Romans' theory can now be uplifted not just to massive IIA using the original ansatz of \cite{Cvetic:1999un}, but also to type IIB. Neglecting isolated examples, since we have worked with a reasonably general ansatz, this work also constitutes a general check of the expectation that non-Abelian T-duality is a symmetry of the equations of motion of type II supergravity.

We have also seen that the correct KK reduction ansatz follows as a result of simply gauging the $S^2$ associated to the $SU(2)$ R-symmetry in the non-Abelian T-dual geometry. Closure of the type IIB field strength $H$ then determines the accompanying $B$-field and the RR sector follows from a requirement that both the original reduction of \cite{Cvetic:1999un} and our new reduction give the same theory in seven-dimensions. We have independently noted that one can perform an $SU(2)$ non-Abelian T-duality transformation following \cite{Itsios:2013wd} to generate the ansatz.  Indeed, if this consistent reduction did not exist, we would be most surprised since it would fly in the face of the conjecture of \cite{Gauntlett:2007ma}. Having identified the expected KK reduction in this paper and through it provided another example, steps towards a proof of this conjecture would be welcome. It is possible that the reduction of the fermions (for example \cite{Bah:2010cu, Bah:2010yt}) may be useful in this regard.  

Using this new connection between Romans' theory and type IIB supergravity we have presented some sample uplifted solutions. Building on the observation that the $AdS_6$ vacuum uplifted to either IIA or IIB is supersymmetric, here we perform similar uplifts for more involved supersymmetric solutions to F(4) gauged supergravity. We begin by uplifting a domain wall solution without $SU(2)$ gauge fields but supported through a non-trivial scalar, before moving onto a supersymmetric $AdS_4 \times H^2$ fixed-point corresponding to a twist of the theory where a $U(1)$ gauge field is excited. Though it is widely assumed that supersymmetry is preserved when one uplifts, here we have taken steps to show that the uplifted solutions are consistent with this expectation. Again the reduction of the Killing spinor equations would help us confirm that the supersymmetric structure is the same.

Finally, one may wonder if the two known reductions from type II supergravity to F(4) gauged supergravity are the whole story?  Certainly we are aware that F(4) gauged supergravity can be coupled to vector multiplets \cite{D'Auria:2000ad}, so one may expect that there is a more general reduction from massive IIA where additional scalars and vectors from the coset $SL(5,\mathbb{R})/SO(5)$ are retained. It would be interesting to address this possibility as it may serve as a stepping stone to the construction of gravity duals where conformal symmetry is broken.

\acknowledgments
We wish to thank P. Karndumri, C. N\'u\~nez, S. Parameswaran and I. Y. Park for correspondence and  Y. Lozano, P. Meessen, D. Rodr\'{i}guez-G\'omez and K. Sfetsos for constructive discussions. In addition, we are grateful to K. Sfetsos and O. Varela for taking the time to critically read a late draft.  J. J. is supported by the National Research Foundation (NRF) of Korea grant funded by the Korea government (MEST) with the grant number 2012-046278 and by the grant number 2005-0049409 through the Center for Quantum Spacetime (CQUeST) of Sogang University. O.K. acknowledges partial support from the Korea Research Foundation Grant funded by the Korean Government (MOEHRD, Basic Research Promotion Fund) (KRF-2007-331-C00109). E. \'O C. is grateful to Ollscoil na h\'Eireainn, M\'a Nuad, where some of this work was performed, and also acknowledges support from the research grants MICINN-09-FPA2009-07122 and MEC-DGI-CSD2007-00042.
\appendix

\section{Type II supergravity EOMs}
\label{sec:IIEOMs}
For completeness here we record the equations of motion of both type IIB supergravity \cite{Schwarz:1983qr} and massive IIA \cite{Romans:1985tz}. We follow the conventions of \cite{Itsios:2012dc}.

\subsection*{Type IIB}
The field content of type IIB supergravity includes a metric $g_{MN}$, a scalar dilaton $\Phi$, an antisymmetric tensor $B$-field, a zero-form $C_0$, a two-form $C_2$ and four-form Ramond potential $C_4$. The corresponding field strengths are
\be
H = d B, \quad F_1 = d C_0, \quad F_3 = d C_2 - C_0 H, \quad F_5 = d C_4 - H \wedge C_2,
\ee
leading to the following Bianchi identities
\be
\label{BianchiIIB}
d H = 0, \quad d F_1= 0, \quad d F_3 = H \wedge F_1, \quad d F_5 = H \wedge F_3.
\ee
The field strength (flux) equations of motions are
\bea
\label{fluxIIBeq1} &&d (e^{-2 \Phi} * H) - F_1 \wedge * F_3 - F_3 \wedge F_5 = 0, \\
\label{fluxIIBeq2} && d * F_1  + H \wedge * F_3 = 0, \\
\label{fluxIIBeq3} && d * F_3 + H \wedge F_5 = 0, \\
\label{fluxIIBeq4} && d * F_5 - H \wedge F_3 =0
\eea
The self-duality condition on $F_5$, i.e. $F_5 = * F_5$, means that (\ref{fluxIIBeq4}) simply reproduces the Bianchi identity.

Finally, the Einstein equation is
\bea
\label{EinsteinIIB} && R_{MN} + 2 \nabla_{M} \nabla_{N} \Phi - \tfrac{1}{4} H^2_{MN} \\
&& \phantom{xxxxxx} = e^{2 \Phi} \left[ \tfrac{1}{2} (F_1^2)_{MN} + \tfrac{1}{4} (F_3^2)_{MN} + \tfrac{1}{96} (F_5^2)_{MN} - \tfrac{1}{4} g_{MN} \left( F_1^2 + \tfrac{1}{6} F_3^2 \right) \right], \nonumber
\eea
and the dilaton satisfies the equation
\be
\label{dileq}
R + 4 \nabla^2 \Phi - 4 (\partial \Phi)^2 - \tfrac{1}{12} H^2 = 0.
\ee

\subsection*{Massive IIA}
The field content of massive IIA supergravity is the same as the above except that the Ramond potentials are now odd-forms, $C_1$ and $C_3$, and the theory has a mass parameter $m$. The field strengths are now
\be
H = d B, \quad F_2 = d C_1 + m B, \quad F_4 = d C_3 - H \wedge C_1 + \tfrac{m}{2} B \wedge B
\ee
with Bianchi identities
\be
\label{BianchiIIA} d H =0, \quad d F_2 = m H, \quad d F_4 = H \wedge F_2.
\ee
The flux equations of motions are then
\bea
\label{fluxIIAeq1} &&d (e^{-2 \Phi} * H) - F_2 \wedge * F_4 - \tfrac{1}{2} F_4 \wedge F_4 =  m * F_2, \\
\label{fluxIIAeq2} && d * F_2  + H \wedge * F_4 = 0, \\
\label{fluxIIAeq3} && d * F_4 + H \wedge F_4 = 0,
\eea
and the Einstein equation becomes
\bea
\label{EinsteinIIA} && R_{MN} + 2 \nabla_{M} \nabla_{N} \Phi - \tfrac{1}{4} H^2_{MN} \\
&& \phantom{xxxxxx} = e^{2 \Phi} \left[ \tfrac{1}{2} (F_2^2)_{MN} + \tfrac{1}{12} (F_4^2)_{MN} - \tfrac{1}{4} g_{MN} \left( \tfrac{1}{2} F_2^2 + \tfrac{1}{24} F_4^2 + m^2 \right) \right]. \nonumber
\eea
As the dilaton equation does not involve the Ramond potentials it is unchanged.

\section{Gauging the $S^2$}
\label{sec:S2gauge}
In this section we give some details about the process through which one may gauge the two-sphere to introduce $SU(2)$ gauge fields. We adopt the usual choice for the metric on $S^2$, $ds^2 = d \theta^2 + \sin^2 \theta d \phi^2$ and proceed to introduce $\mu^i$, $i=1, 2, 3$ satisfying $\mu^i \mu^i =1$ which parametrise the two-sphere.  Given our choice of the metric, the three Killing vectors on the $S^2$ are
\bea
K_1 &=& - \cos \phi \partial_{\theta} + \cot \theta \sin \phi \partial_{\phi}, \nn
K_2 &=& \sin \phi \partial_{\theta} + \cot \theta \cos \phi \partial_{\phi}, \nn
K_3 &=& \partial_{\phi}.
\eea
One can check that these Killing vectors satisfy the commutation relations of the $SU(2)$ Lie algebra, i.e. $[ K_i, K_j] = \e_{ijk} K_k$. We now introduce the usual frame for the $S^2$
\be
e^{\theta} = d \theta, \quad e^{\phi} = \sin \theta d \phi,
\ee
allowing us to define the dual vectors
\be
e_{\theta} = \partial_{\theta}, \quad e_{\phi} = \frac{1}{\sin \theta} \partial_{\phi}.
\ee
The Killing vectors above are written with respect to coordinates, but we can rewrite them in terms of the dual vectors as
\bea
K^{\theta}_1 &=& - \cos \phi, \quad K^{\phi}_1 = \cos \theta \sin \phi, \nn
K^{\theta}_2 &=& \sin \phi, \quad K^{\phi}_2 = \cos \theta \cos \phi, \nn
K^{\phi}_3 &=& \sin \theta.
\eea
One can check that these satisfy the following relationships:
\be
K^{i a} K^j_{a} = \delta^{ij} - \mu^i \mu^j, \quad K_{i}^a K_i^b = \delta^{ab}.
\ee
We can now define the metric on the original $S^2$ as
\be
ds^2(S^2) = d \mu^i d \mu^i,
\ee
where $ d \mu^i = \e^{ab} K^i_{b} e^{b}$ \footnote{We take $\e^{\theta \phi} = 1$.}. This then leads to an explicit representation for $d \mu^i$ and $\mu^i$:
\be
\mu^1 =\sin \theta \sin \phi,  \quad \mu^2 = \sin \theta \cos \phi, \quad \mu^3 = - \cos \theta.
\ee
One can confirm that $\frac{1}{2} \e_{ijk} \mu^i d \mu^j \wedge d \mu^k = \vol(S^2)$. We are now in a position to introduce a gauging of the $S^2$ through
\be
D \mu^i = \e^{ab} K^i_{b} (e_a - K^k_{a} A^k) = d \mu^i - \e_{ijk} \mu^j A^k,
\ee
where we have introduced $SU(2)$ gauge fields $A^k$. It is useful to document the following:
\bea
d (\tfrac{1}{2}  \e_{ijk} \mu^i D \mu^j \wedge D \mu^k) &=& D \mu^i \wedge \left[ d A^i + \tfrac{1}{2} \e^{ijk} A^j \wedge A^k \right], \nn
&=& D \mu^i \wedge {F}^i.
\eea

\section{Non-Abelian T-duality}
\label{sec:NAT}
In this section we show that a non-Abelian T-duality transformation of the NS sector of the original ansatz (\ref{KKansatz}) leads to the T-dual NS sector quoted in the text on the nose. Recall that the NS sector of our original massive IIA space-time is of the following form
\bea
ds^2 &=& G_{\mu \nu} dx^{\mu} dx^{\nu} + 2 G_{\mu i} dx^{\mu} \s^i + g_{ij} \s^i \s^j, \nn
B &=& \tfrac{1}{2} B_{\mu \nu} dx^{\mu} \wedge dx^{\nu},
\eea
where $\s^i$ denote the left-invariant one-forms as before and of course, we have an additional dilaton. Comparison with (\ref{KKansatz}) reveals that
\be
G_{\mu \nu} = g_{\mu \nu} + e^{2A} A^i_{\mu} A^i_{\nu}, \quad  g_{ij} = e^{2A} \delta_{ij}, \quad G_{\mu i } = - e^{2A} A^i_{\mu}, \quad B_{\mu i} = 0,
\ee
where $g_{\mu \nu}$ denotes the metric on $M_7$.

As explained in detail in \cite{Sfetsos:2010uq, Itsios:2013wd}, a generic $SU(2)$ transformation depends on a matrix of the form
\be
M_{ij} = e^{2A} \delta_{ij} - \e_{ijk} x^k,
\ee
where $x^k$ is a Lagrange multiplier, or alternatively a dual coordinate once one does the $SU(2)$ transformation, and the minus sign appears above as we are doing a transformation with respect to left-invariant one-forms. The inverse matrix is then
\be
M_{ij}^{-1} = \frac{1}{e^{2 A} (r^2 + e^{4 A})} \left( \begin{array}{ccc} e^{4 A} + x_1^2 & x_1 x_2 + e^{2A} x_3 & x_1 x_3 - e^{2A} x_2 \\ x_1 x_2 - e^{2A} x_3 & e^{4 A} + x_2^2 & x_2 x_3 + e^{2A} x_1\\ x_1 x_3 + e^{2A} x_2 & x_2 x_3 - e^{2A} x_1& e^{4 A} + x_3^2 \end{array} \right),
\ee
where we have introduced a natural radial coordinate, $r^2 = x_i x^i $.

Then, defining the following
\be
Q_{\mu \nu} = G_{\mu \nu} + B_{\mu \nu}, \quad Q_{\mu i} = G_{\mu i} + B_{\mu i}, \quad Q_{i \mu} = G_{i \mu} + B_{i \mu},
\ee
the non-Abelian T-dual can be read off from
\be
\label{nonabt}
\tilde{Q}_{\mu \nu} = Q_{\mu \nu} - Q_{\mu i} M_{ij}^{-1} Q_{j \nu}, \quad \tilde{Q}_{\mu i} = Q_{\mu j} M^{-1}_{ji}, \quad \tilde{Q}_{i \mu} = - M_{ij}^{-1} Q_{j \mu}.
\ee
This leads to the metric (\ref{iibmet}) and the $B$-field (\ref{iibB}) quoted in the text once one rewrites $x^i = r \mu^i$ in terms of the constrained coordinates on the $S^2$.

\subsection*{RR fluxes}
To complete the ansatz we have to perform the accompanying transformation for the RR fluxes. Here we simply sketch the calculation and refer the reader to \cite{Itsios:2013wd} for further details. After constructing the flux bispinor for the original solution $P$, one operates with $\Omega^{-1}$ to get the T-dual bispinor $\hat{P}$ and then extracts the various components of the fluxes:  
\be
\hat{P} = P \cdot \Omega^{-1}  = P \cdot \G_{11} \frac{e^{2A} \G_{789} + x_i \G^{i}}{\sqrt{r^2 + e^{4A}}}, 
\ee
where $i = 7, 8, 9$ denote $S^3$ directions and for concreteness we take $\G_{11} = \G_{0123456789}$. In defining the bispinors we use 
\be
P = \frac{e^{\Phi}}{2} \sum_{n=0}^5 \slashed{F}_{2n}, \quad \hat{P} = \frac{e^{\tilde{\Phi}}}{2} \sum_{n=0}^4 \slashed{\tilde{F}}_{2n+1}, 
\ee
where $\slashed{F} = \frac{1}{p!} F_{\mu_1\dots \mu_p} \G^{\mu_1 \dots \mu_p}$ for a $p$-form flux. 
In reconstructing the T-dual forms one has to make use of the appropriate frame \cite{Itsios:2013wd}
\be
\hat{e}^{i} = e^{-A} \mu^i dr + \frac{e^{-A}}{r^2 + e^{4A}} \left[ r e^{4A} D \mu^i - r^2 e^{2A} \e_{ijk} \mu^j D \mu^k \right]. 
\ee
In addition, we find the following relations useful 
\be
x_i \hat{e}^i = e^{-A} r dr, \quad e^{A} \hat{e}^i + e^{-A} \e_{ijk} x^j \hat{e}^k = \mu^i dr + r D \mu^i. 
\ee
With a little care one can show that the RR fluxes for type IIB presented in the text are simply the non-Abelian T-dual of the massive IIA fluxes using the above prescription for the transformation. 

\section{Details of some calculations}
\subsection*{Massive IIA reduction}
Here we record some useful expressions. The Hodge duals of the fluxes are
\bea
\label{IIAhodge}
*\;F_2 &=& e^{3 A} (*_7 G_2) \wedge h_1 \wedge h_2 \wedge h_3, \\ \nonumber
* \;H  &=&   e^{3 A} (*_7 H) \wedge h_1 \wedge h_2 \wedge h_3, \\ \nonumber
* \;F_4 &=&  e^{3 A} (*_7 G_4) \wedge h_1 \wedge h_2 \wedge h_3 + e^{-3A} (*_7 G_1) \nn
&-& \tfrac{1}{2} e^{A} \e_{ijk} (*_7 H^i_3) \wedge h_j \wedge h_k + e^{-A} (*_7 H_2^i) \wedge h_i.
\eea

Making use of the orthonormal frame (\ref{frame1}) one can work out the spin connection from derivatives of the vielbein. One first determines $c^{M}_{~NP}$ from
\be
d e^{M} = - c^{M}_{~NP} e^{N} \wedge e^{P}
\ee
and then calculates $\omega_{MNP}$ (lowering appropriate indices)
\be
\omega_{MNP} = \tfrac{1}{2} \left( c_{MNP} + c_{NMP} - c_{PMN}\right).
\ee
The spin connection one-form is then $\omega^{N}_{~P} = \omega_{M~P}^{~N}  e^{M}$. We can thus determine the spin connection for the above orthonormal frame (\ref{frame1}) and get
\bea
\label{spin1}
\omega^{1}_{~2} &=& - A^3_{\mu} e^{\mu} - \tfrac{1}{2} e^{-A} e^3, \nn
\omega^{2}_{~3} &=& -  A^1_{\mu} e^{\mu} - \tfrac{1}{2} e^{-A} e^1, \nn
\omega^{3}_{~1} &=& -  A^2_{\mu} e^{\mu} - \tfrac{1}{2} e^{-A} e^2,  \nn
\omega^{1}_{~\mu} &=& \partial_{\mu} A e^1 - \tfrac{1}{2} e^{A} F^1_{\mu \nu} e^{\nu}, \nn
\omega^{2}_{~\mu} &=& \partial_{\mu} A e^2 - \tfrac{1}{2} e^{A} F^2_{\mu \nu} e^{\nu} , \nn
\omega^{3}_{~\mu} &=& \partial_{\mu} A e^3 - \tfrac{1}{2} e^{A} F^3_{\mu \nu} e^{\nu}, \nn
\omega^{\mu}_{~\nu} &=& \bar{\omega}^{\mu}_{~\nu} + \tfrac{1}{2} e^{A} F^{i \mu}_{~~\nu} e^i,
\eea
where $\bar{\omega}$ denotes the spin connection purely on $M_7$. For consistency one can check these satisfy $d e^{M} + \omega^{M}_{~N} e^{N} = 0$. In calculating the Ricci tensor it is good to use
\bea
d e^{\mu} &=& - \bar{\omega}^{\mu}_{~\nu} e^{\nu}, \nn
d e^i &=& \partial_{\mu} A e^{\mu i} - e^{A} F^i - \e_{ijk} \left( \tfrac{1}{2} e^{-A} e^{jk} + e^{j} A^{k} \right).
\eea

\subsection*{IIB reduction}
In deriving the equations of motion we have made use of the following Hodge duals
\bea
\label{IIBhodge}
{*} F_1 &=& \frac{e^{A} r^2}{r^2 + e^{4A}}  \left[ - *_7 G_1 \wedge dr  -m r e^{2A} \vol(M_7) \right] \wedge \vol(\tilde{S}^2) , \nn
{*} F_3 &=& \frac{e^{3A} r^2}{r^2 + e^{4A}} \left[ - e^{A}  G_4 dr {-} r *_7 G_2 \right] \wedge \vol(\tilde{S}^2) \nn
& +& r e^{-3A} *_7 G_1 \wedge dr - m e^{3 A} \vol(M_7) - \frac{r^3 e^{A}}{r^2 + e^{4A}} \mu^i *_7 H_3^i \wedge dr \wedge \vol(\tilde{S}^2) \nn
&+& \mu^i  *_7 H_2^i \frac{r^2 e^{3 A}}{r^2 + e^{4 A}} \vol(\tilde{S}^2)+ r e^{-A} *_7 H_2^i \wedge dr \wedge \e_{ijk} \mu^j D \mu^k .
\eea
As we are now in type IIB, the five-form flux is self-dual, $*F_5 = F_5$, so we do not need the Hodge dual for $F_5$. For certain terms it is good to use the identity
\be
*_2 D\mu^i = - \e_{ijk} \mu^j D \mu^k,
\ee
where $*_2$ refers to Hodge duality on the $S^2$.

Here we record various derivatives of the vielbein (\ref{frameiib}) presented in the text
\bea
\label{derivatives}
d e^{\mu} &=& - \bar{\omega}^{\mu}_{~\nu} e^{\nu}, \nn
d e^r &=& - \partial_{\mu} A e^{\mu r}, \nn
d e^1 &=& \frac{(r^2-e^{4A})}{(r^2 + e^{4A}) } \partial_{\mu} A e^{\mu 1} + \frac{e^{5 A}}{r (r^2 + e^{4 A})} e^{r1} - \frac{1}{\sin \theta}  e^2 \left( \sin \phi A^1 + \cos \phi A^2 \right) \nn &-& \frac{r e^{A}}{\sqrt{r^2 + e^{4 A}}} K^i_{\theta} F^i, \nn
d e^2 &=& \frac{(r^2-e^{4A})}{(r^2 + e^{4A}) } \partial_{\mu} A e^{\mu 2} + \frac{e^{5 A}}{r (r^2 + e^{4 A})} e^{r 2} + \frac{1}{\sin \theta} e^1 \left(  \sin \phi A^1 + \cos \phi A^2 \right) \nn &-& \frac{r e^{A}}{\sqrt{r^2 + e^{4 A}}} K^i_{\phi} F^i + \cot \theta \frac{\sqrt{r^2 + e^{4 A}}}{r e^A} e^{12}.
\eea
Making use of these above expressions, one can determine the spin connection:
\bea
\label{spiniib}
\omega^{\mu}_{~\nu} &=& \bar{\omega}^{\mu}_{~\nu} + \frac{1}{2} \frac{r e^{A}}{\sqrt{r^2 + e^{4A}}} K^{i}_{a} F^{i \mu}_{~~\nu} e^{a}, \nn
\omega^{r}_{~\mu} &=& - \partial_{\mu} A e^r, \nn
\omega^{a}_{~r} &=& \frac{1}{r} \frac{e^{5A}}{(r^2 + e^{4 A})} e^{a}, \nn
\omega^{a}_{~\mu} &=& \frac{(r^2 - e^{4A})}{(r^2 + e^{4A})} \partial_{\mu} A e^{a} + \frac{1}{2} \frac{r e^{A}}{\sqrt{r^2 + e^{4A}}} K^{i}_{a} F^{i}_{~ \rho \mu} e^{\rho}, \nn
\omega^{1}_{~2} &=& - \frac{\sqrt{r^2 + e^{4A}}}{r e^{A}} \cot \theta e^2 - \frac{1}{\sin \theta} \left( \sin \phi A^1 + \cos \phi A^2 \right).
\eea
where we have used $a=1, 2$.

\subsection*{Miscellaneous}
Here we present some details for the calculation of $\nabla_{i} \nabla_{\mu} \Phi$. By definition this is
\be
\label{covdiff}
\nabla_{i} \nabla_{\mu} \Phi = \partial_{i} \partial_{\mu} \Phi - \G^{\nu}_{i \mu} \partial_{\nu} \Phi,
\ee
where the $i = 1, 2, 3$ index refers to orthonormal frame and since $\Phi$ only depends on the coordinates on the $M_7$ the first term disappears so we only need determine the second term. Specialising to the case where the $SU(2)$ gauge fields are truncated to retain a $U(1)$, we can introduce the vielbein
\bea
e^{1} &=& e^{A} d \theta, \nn
e^{2} &=& e^{A} \sin \theta d \phi, \nn
e^{3} &=& e^{A} ( d \psi + \cos \theta d \phi - A_{\mu} \bar{e}^{\mu} ), \nn
e^{\mu} &=& \bar{e}^{\mu},
\eea
and invert it to get the inverse vielbein
\bea
e^{1} &=& e^{-A} \partial_{\theta}, \nn
e^{2} &=& e^{-A} \left( \frac{1}{\sin \theta} \partial_{\phi} - \frac{\cos \theta}{\sin \theta} \partial_{\psi} \right), \nn
e^{3} &=& e^{-A} \partial_{\psi}, \nn
e^{\mu} &=& A^{\mu} \partial_{\psi} + \bar{\partial}_{\mu}.
\eea
We clearly see from these that the first term in (\ref{covdiff}) disappears. Now, as $\G^{M}_{PQ} = \frac{1}{2} g^{MN} (g_{PN, Q} + g_{QN,P} - g_{PQ,N})$, where $g_{MN}$ is the ten-dimensional metric, we need to determine the inverse metric. Doing so, we find the following matrix
\be
g^{MN} = \left( \begin{array}{cccc} e^{-2A} & 0 & 0 & 0 \\ 0 & e^{-2A} \frac{1}{\sin^2 \theta} & - e^{-2 A} \frac{\cos \theta}{\sin^2 \theta} & 0 \\  0 & - e^{-2 A} \frac{\cos \theta}{\sin^2 \theta} & g^{\mu \nu} A_{\mu} A_{\nu} + e^{-2A} \frac{1}{\sin^2 \theta} & g^{\mu \nu} A_{\nu} \\ 0 & 0 & g^{\mu \nu} A_{\nu} & g^{\mu \nu} \end{array} \right).
\ee
Once we have the inverse metric and the inverse vielbein we can calculate the Christoffel symbols in orthonormal frame. One finds that
\be
\G^{\nu}_{3 \mu} \partial_{\nu} \Phi = e^{A} F^{\nu}_{~\mu} \partial_{\nu} \Phi.
\ee
is non-zero. Though more involved, the generalisation to include the $SU(2)$ gauge fields is straightforward and leads to expression on the LHS of (\ref{iibEcross}).

\end{document}